\documentclass[fleqn,usenatbib]{mnras}

\usepackage{newtxtext,newtxmath}
\usepackage[T1]{fontenc}
\usepackage{float}

\DeclareRobustCommand{\VAN}[3]{#2}
\let\VANthebibliography\thebibliography
\def\thebibliography{\DeclareRobustCommand{\VAN}[3]{##3}\VANthebibliography}

\usepackage{graphicx}	% Including figure files
\usepackage{amsmath}	% Advanced maths commands
\usepackage{multirow}

\usepackage{soul}
\usepackage{orcidlink}

\usepackage{hyperref}

\newcommand{\ndet}{462}
\newcommand{\nlos}{2,714}
\newcommand{\ncnm}{691}
\newcommand{\nbiglos}{264}
\newcommand{\ntsbiglos}{117}

\newcommand{\ntsbigcnm}{400}

\newcommand{\Tc}{$T_\mathrm{c}$}

\newcommand{\kms}{km s$^{-1}$}

\def\13co{$^{13}$CO}
\def\aj{AJ}

\def\aap{A\&A}
\def\apj{ApJ}
\def\apjs{ApJS}
\def\pasa{PASA}

\def\av{$A_\mathrm{V}$}
\def\c18o{C$^{18}$O}

\def\cm2{cm$^{-2}$}
\def\cm3{cm$^{-3}$}

\def\ebv{$E(B{-}V)$}
\def\FCNM{$f_\mathrm{CNM}$}
\def\RHI{$\mathcal{R_{\mathrm{H{\sc I}}}}$}
\def\h2{H$_2$}
\def\hi{H{\sc i}}

\def\kms{km s$^{-1}$}
\def\k0{$\kappa_{0}$}

\def\nat{Nature}
\def\n2h{N$_2$H$^+$}
\def\NHI{$N_\mathrm{H{\sc I}}$}
\def\NHICNM{$N_\mathrm{H{\sc I},CNM}$}

\def\nhiUnit{$\times\ 10^{20}$ cm$^{-2}$}
\def\NHIthin{$N^{*}_\mathrm{HI}$}

\def\s{s$^{-1}$}
\def\s353{$\sigma_{353}$}

\def\rv{$R_\mathrm{V}$}

\def\taupeak{$\tau_{\mathrm{peak}}$}

\def\Tbv{$T_{\mathrm{b}}(v)$}
\def\TBpeak{$T_{\mathrm{b,peak}}$}
\def\Tbg{$T_{\mathrm{bg}}$}
\def\Tc{$T_{\mathrm{c}}$}

\def\TD{$T_{\mathrm{D}}$}
\def\Tk{$T_{\mathrm{k}}$}
\def\Tkmax{$T_{\mathrm{k,max}}$}
\def\Ts{$T_{\mathrm{s}}$}
\def\t353{$\tau_{353}$}

\title[Magellanic Cloud foreground \hi\ absorption with ASKAP]{Local \hi\ Absorption towards the Magellanic Cloud foreground using ASKAP}

\author[H. Nguyen et al.]{Hiep Nguyen\orcidlink{https://orcid.org/0000-0002-2712-4156}$^{1}$\thanks{E-mail: \href{mailto:vanhiep.nguyen@anu.edu.au}{vanhiep.nguyen@anu.edu.au}},
N.~M. McClure-Griffiths\orcidlink{https://orcid.org/0000-0003-2730-957X}$^{1}$,
James Dempsey\orcidlink{https://orcid.org/0000-0002-4899-4169}$^{1,2}$,
John M. Dickey\orcidlink{https://orcid.org/0000-0002-6300-7459}$^{3}$,
Min-Young Lee\orcidlink{https://orcid.org/0000-0002-9888-0784}$^{4}$, \newauthor
Callum Lynn\orcidlink{https://orcid.org/0000-0001-6846-5347}$^{1}$,
Claire Murray\orcidlink{https://orcid.org/0000-0002-7743-8129}$^{5,6}$,
Sne\v{z}ana Stanimirovi\'{c}\orcidlink{https://orcid.org/0000-0002-3418-7817}$^{7}$,
Michael P. Busch\orcidlink{0000-0003-4961-6511}$^{8}$,
Susan E. Clark\orcidlink{https://orcid.org/0000-0002-7633-3376}$^{9,10}$,
\newauthor
J. R.  Dawson\orcidlink{https://orcid.org/0000-0003-0235-3347}$^{11,12}$,
Helga Dénes\orcidlink{https://orcid.org/0000-0002-9214-8613}$^{13}$,
Steven Gibson\orcidlink{https://orcid.org/0000-0002-1495-760X}$^{14}$,
Katherine Jameson\orcidlink{https://orcid.org/0000-0001-7105-0994}$^{15}$,
Gilles Joncas\orcidlink{https://orcid.org/0000-0001-7462-4818}$^{16}$,
\newauthor
Ian Kemp\orcidlink{https://orcid.org/0000-0002-6637-9987}$^{17}$,
Denis Leahy\orcidlink{https://orcid.org/0000-0002-4814-958X}$^{18}$,
Yik Ki Ma\orcidlink{https://orcid.org/0000-0003-0742-2006}$^{1}$,
Antoine Marchal\orcidlink{https://orcid.org/0000-0002-5501-232X}$^{1}$,
Marc-Antoine Miville-Desch\^{e}nes\orcidlink{https://orcid.org/0000-0002-7351-6062}$^{19}$,
\newauthor
Nickolas M. Pingel\orcidlink{https://orcid.org/0000-0001-9504-7386}$^{7}$,
Amit Seta\orcidlink{https://orcid.org/0000-0001-9708-0286}$^{1}$, 
Juan D. Soler\orcidlink{https://orcid.org/0000-0002-0294-4465}$^{20}$, 
Jacco Th. van Loon\orcidlink{https://orcid.org/0000-0002-1272-3017}$^{21}$
\\
\\
$^{1}$Research School of Astronomy and Astrophysics, The Australian National University, Canberra, ACT 2611, Australia\\
$^{2}$CSIRO Information Management and Technology, GPO Box 1700 Canberra, ACT 2601, Australia\\
$^{3}$School of Natural Sciences, Private Bag 37, University of Tasmania, Hobart, TAS, 7001, Australia\\
$^{4}$Korea Astronomy and Space Science Institute, 776, Daedeokdae-ro, Yuseong-gu Daejeon 34055, Republic of Korea\\
$^{5}$Department of Physics \& Astronomy, Johns Hopkins University, 3400 N. Charles Street, Baltimore, MD 21218, USA\\
$^{6}$Space Telescope Science Institute, 3700 San Martin Drive, Baltimore, MD 21218, USA\\
$^{7}$Department of Astronomy, University of Wisconsin, Madison, WI 53706-15821, USA\\
$^{8}$Department of Astronomy \& Astrophysics, University of California, San Diego, 9500 Gilman Drive, La Jolla, CA 92093, USA\\
$^{9}$Department of Physics, Stanford University, 382 Via Pueblo Mall, Stanford, CA 94305, USA\\
$^{10}$Kavli Institute for Particle Astrophysics \& Cosmology, P.O. Box 2450, Stanford University, Stanford, CA 94305, USA\\
$^{11}$Department of Physics and Astronomy and MQ Research Centre in Astronomy, Astrophysics, and Astrophotonics, Macquarie University, NSW 2109, Australia\\
$^{12}$Australia Telescope National Facility, CSIRO Space and Astronomy, PO Box 76, Epping NSW 1710, Australia\\
$^{13}$School of Physical Sciences and Nanotechnology, Yachay Tech University, Hacienda San Jos\'e S/N, 100119, Urcuqu\'i, Ecuador\\
$^{14}$Department of Physics and Astronomy, Western Kentucky University, Bowling Green, KY 42101, USA\\
$^{15}$Caltech Owens Valley Radio Observatory, Pasadena, CA 91125, USA\\
$^{16}$D\'{e}partement de physique, de g\'{e}nie physique et d'optique, Universit\'{e} Laval, Pavillon Alexandre-Vachon 1045, Avenue de la M\'{e}decine, Qu\'{e}bec City, Canada\\ and Centre de recherche en astrophysique du Qu\'{e}bec\\
$^{17}$International Centre for Radio Astronomy Research (ICRAR), Curtin University, Bentley, WA 6102, Australia\\
$^{18}$Department of Physics and Astronomy, University of Calgary, Calgary, AB T2N 1N4, Canada\\
$^{19}$AIM, CEA, CNRS, Université Paris-Saclay, Universit\'{e} Paris Diderot, Sorbonne Paris Cit\'{e}, F-91191 Gif-sur-Yvette, France\\
$^{20}$Istituto di Astrofisica e Planetologia Spaziali (IAPS). INAF. Via Fosso del Cavaliere 100, 00133 Roma, Italy\\
$^{21}$Lennard-Jones Laboratories, Keele University, ST5 5BG, UK
}

\date{Accepted XXX. Received YYY; in original form ZZZ}

\pubyear{2024}

\begin{document}
\label{firstpage}
\pagerange{\pageref{firstpage}--\pageref{lastpage}}
\maketitle

\begin{abstract}
We present the largest Galactic neutral hydrogen \hi\ absorption survey to date, utilizing the Australian SKA Pathfinder Telescope at an unprecedented spatial resolution of 30$^{\prime \prime}$. This survey, GASKAP-\hi, unbiasedly targets \nlos\ continuum background sources over 250 square degrees in the direction of the Magellanic Clouds, a significant increase compared to a total of 373 sources observed by previous Galactic absorption surveys across the entire Milky Way. We aim to investigate the physical properties of cold (CNM) and warm (WNM) neutral atomic gas in the Milky Way foreground, characterized by two prominent filaments at high Galactic latitudes (between $-45^{\circ}$ and $-25^{\circ}$). We detected strong \hi\ absorption along \ndet\ lines of sight above the 3$\sigma$ threshold, achieving an absorption detection rate of 17\%. GASKAP-\hi's unprecedented angular resolution allows for simultaneous absorption and emission measurements to sample almost the same gas clouds along a line of sight. A joint Gaussian decomposition is then applied to absorption-emission spectra to provide direct estimates of \hi\ optical depths, temperatures, and column densities for the CNM and WNM components. The thermal properties of CNM components are consistent with those previously observed along a wide range of Solar neighbourhood environments, indicating that cold \hi\ properties are widely prevalent throughout the local interstellar medium. Across our region of interest, CNM accounts for $\sim$30\% of the total \hi\ gas, with the CNM fraction increasing with column density toward the two filaments. Our analysis reveals an anti-correlation between CNM temperature and its optical depth, which implies that CNM with lower optical depth leads to a higher temperature.
\end{abstract}


%%%%%%%%%%%%%%%%%%%%%%%%%%%%%%%%%%%%%%%%%%%%%%%%%%

%%%%%%%%%%%%%%%%% BODY OF PAPER %%%%%%%%%%%%%%%%%%++

\section{Introduction}
    Neutral atomic hydrogen (\hi), the most abundant gas in the interstellar medium (ISM), plays a crucial role in the evolution and dynamics of the ISM in galaxies. It provides the initial material for the formation of stars, serves as the building block of molecular clouds, influences the dynamics of the ISM, acts as a cooling agent and as a source of radiation shielding, and participates in the feedback mechanism that regulates star formation as well as galaxy evolution. The study of \hi\ gas properties is essential for understanding the structure and composition of the ISM in galaxies. In particular, the characterization of the cold neutral medium (CNM) and warm neutral medium (WNM) is of great interest, as they represent distinct phases with different physical properties \citep[][]{Field1969,Wolfire1995,Wolfire2003,Heiles2003b,Tielens2005,Cox2005}.

The \hi\ gas exists in multiple phases, which are basically characterized by different temperatures and densities: two thermally stable phases, CNM and WNM, with kinetic temperature and density of (60--260 K, 7--70 \cm3) and (5000--8300 K, 0.2--0.9 \cm3), respectively, for typical Solar metallicity \citep{Field1969,McKee1977,Wolfire2003}. Additionally, a thermally unstable phase (or unstable neutral medium, UNM) co-exists, with intermediate temperature and density. The UNM is thought to arise from perturbations caused by non-thermal processes such as supernova shocks, stellar winds, and turbulence \citep[][]{Heiles2003a,Audit2005,Roy2013, Murray2015, Murray2018,Kalberla2018,Marchal2019,Seta2022,Bhattacharjee2024}. The WNM contributes roughly $\sim$50\% of the mass of \hi\ in the ISM, the CNM consists of 30\%, and the remaining 20\% is the contribution of the UNM \citep{Murray2015,Murray2018,McClure-Griffiths2023}.

On the journey from atomic gas to stars, the \hi\ from the warm phase needs to cool down and settles itself into the cold phase. Gravity then causes the cold gas to collapse and form dense regions, known as molecular clouds (T $\sim$ 10 K), where the density becomes high and the temperature is low enough for hydrogen atoms to form molecules. As the density increases further, these molecular clouds collapse and form protostars, which eventually evolve into fully-formed stars. Since the CNM is considered a bridge between warm \hi\ gas and molecular clouds, its mass fraction (\FCNM) is a key parameter for understanding the transition from atomic gas to molecular clouds.

In practice, the combination of 21-cm emission and absorption along a line-of-sight provides the most direct approach to estimating the physical properties of the atomic ISM. Such observational pairs allow us to infer optical depths, temperatures (either spin temperatures or upper limits to kinetic temperature), and column densities -- the key parameters for observationally distinguishing the different phases of \hi\ gas including CNM, UNM, and WNM, as well as constraining their fractions. The column density under the optically-thin assumption ($N^{*}_\mathrm{HI}$) is proportional to the \hi\ brightness temperature, hence, can be readily obtained from observed emission profiles. However, this assumption may miss a significant amount of gas mass, because the emission includes not only contributions from warm, optically-thin gas but also from cold, optically-thick gas. In the case when absorption data are not available, we alternatively have to apply some kind of opacity correction to the available emission data.

In this paper, we focus on the \hi\ clouds at high Galactic latitudes in the Milky Way foreground towards the Magellanic Clouds (MCs), $b$ $\sim$ ($-$45, $-$25)$^{\circ}$. This region is characterized by the intersection of two distinct filaments, composed of gas and dust, as illustrated in Figure \ref{fig:all_src_locations}, Reticulum in the vertical direction (near the Reticulum constellation) and Hydrus in the horizontal direction (near the Hydrus constellation) \citep[see][]{Youssef2024}. These filaments are of particular interest as they appear to reside near the edge of the Local Bubble, a low-density gas-filled region likely formed by stellar feedback \citep{Berkhuijsen1971, Cox1987, McKee1998, Zucker2022}, also being situated at slightly different distances ($\sim$200--300 pc) from the Sun \citep{ONeill2024,Erceg2024}. In order to probe the \hi\ gas in the MC foreground, we utilize the Galactic Australian Square Kilometre Array Pathfinder Survey at an unprecedented angular resolution of 30$^{\prime\prime}$ \citep[GASKAP--\hi;][]{Dickey2013,Pingel2022}. GASKAP-\hi’s large field of view ($5 \times 5$ square degrees) facilitates simultaneous measurements of emission and absorption towards \nlos\ background continuum sources across a 250 square degree area. The high absorption measurement density (11 per square degree) provides a dense grid of \hi\ optical depth, a key observable for constraining the properties of the cold neutral ISM. With high angular resolution, emission can be observed close on the sky to the absorption; therefore they both sample similar gas parcels along a line of sight and at a similar range of scales. In this study, by applying the multiple Gaussian decomposition method introduced by \cite{Heiles2003a} to the emission-absorption pairs, we extract the properties of cold and warm \hi\ gas: the optical depth, spin temperature, maximum kinetic (Doppler) temperature, central velocity, cold gas fraction, CNM and WNM column densities, and the total \hi\ column densities along a line-of-sight. While our primary objective is to examine \hi\ physical properties in this Galactic Cirrus region based on absorption-detected lines of sight, a study of the spatial properties and distribution of cold and unstable \hi\ gas has been carried out by \cite{Lynn2024} using information from stacked non-detection spectra.
Additionally, a detailed analysis of the \hi\ filamentary structure is being conducted in an accompanying paper by Lynn et al. 2024b (in prep).

% Tielens1987ASS
Additionally, we investigate the relationship between the \hi\ gas and interstellar dust. Dust grains are ubiquitous in the ISM, and play an important role in shaping the properties of the gas by providing a surface for gas-phase reactions and cooling \citep{Hollenbach1971,LeitchDevlin1984,Cazaux2010}. The gas-dust relationship is complex and depends on the local physical conditions of the gas and the dust. While gas and dust are generally presumed to be well-mixed in the ISM, leading to a general linear correlation between the two (between dust optical depth, dust reddening and gas column density), recent studies have illuminated deviations, including excesses in dust reddening \ebv\ and dust optical depth in regions with high gas column densities. These excesses are generally linked either to the presence of dark gas (either optically-thick \hi, or CO-dark \h2, or both), or changes in the properties of dust grains \citep{PLC2014,PLC2015,Lenz2017,Remy2017,Okamoto2017,Nguyen2018}. By comparing the derived properties of the \hi\ gas (column densities and cold gas fraction) with dust extinction and dust emission, we aim to understand the connection between the gaseous and dusty components of the high Galactic latitude ISM. 

In the context of this paper, we will consistently adopt the following definitions: ``WNM'' refers to \hi\ components that are exclusively detected in emission, while ``CNM'' refers to components detected in absorption.  This approach is reasonable given the moderate optical depth sensitivity ($\sim$0.04 per 0.977 \kms\ velocity channel) of GASKAP-\hi\ observations; in contrast, high optical depth sensitivity observations can even detect WNM in absorption \citep[e.g,][]{Murray2018}.

In this work, we present the GASKAP-\hi\ absorption measurements from the  Pilot Phase II Magellanic Cloud \hi\ foreground observations \citep[see][]{Pingel2022} to explore the properties of \hi\ gas in the Magellanic foreground in an unbiased way. Section \ref{sec:observations} outlines the GASKAP pilot II observations of the MC foreground, detailing the process of extracting emission and absorption spectra around the continuum background sources. Section \ref{subsec:auxiliary_data} lists the auxiliary datasets used in this work. We summarize the method of Gaussian decomposition of emission-absorption pairs in Section \ref{subsec:gaussian_fitting}. Section \ref{sec:component_properties} describes the properties of CNM and WNM components resulting from our fitting, followed by a description of the \hi\ gas characteristics along lines of sight in Section \ref{subsec:los_properties}. In section \ref{sec:hi_dust}, we discuss the relation between gas and dust in the MC foreground. Finally, Section \ref{sec:conclusions} provides a summary of our findings.

% \section{Introduction}

\section{Observations}
    \label{sec:observations}

The GASKAP-\hi\ Pilot II survey targeted ten fields (six toward the LMC, three toward the Bridge, and one toward the SMC) in the direction of the Magellanic System using the Australian Square Kilometre Array Pathfinder (ASKAP, e.g. \citealt{Johnston2007,Hotan2021}) interferometer from June 2020 to March 2022. Each field was observed for 10 hours using the standard GASKAP-\hi\ observing configurations \citep[see][]{Pingel2022}. The survey has been designed to fully resolve the spectral line widths observed in cold \hi\ with a required frequency resolution of $\nu \leq$ 5 kHz. The closepack-36 phased-array feed (PAF) footprint observation scheme was employed with a pitch of 0.9 deg and three interleaves to ensure even coverage across the 25 square degree field.

The GASKAP-\hi\ survey utilized the ``zoom-16'' mode to achieve a spectral resolution of $\nu$ = 1.15 kHz, giving a velocity resolution of $v \sim$ 0.244 km s$^{-1}$ over a total bandwidth of 18 MHz centered on 1419.81 MHz in the kinematic local standard of rest (LSRK) reference frame spanning 15,552 fine channels. We however inclusively select only 2048 channels from 7,887 to 9,934, equivalent to a velocity range of ($-100$, $+400$) km s$^{-1}$ in the LSRK reference frame, which covers \hi\ emission from the Magellanic System and Milky Way, to reduce the total amount of data that needs processing.

The emission survey data were calibrated and reduced using the WSClean software package \citep{offringa-wsclean-2017,vandertol-2018}, which applies standard techniques (such as flagging, calibration, and imaging) with the configuration optimized for wide-field emission \citep{Hotan2021}. Our analysis focuses on the foreground \hi\ gas in the Milky Way within the velocity range from $-50$ to $+50$ km s$^{-1}$, where we can detect emission/absorption features above the 3$\sigma$ threshold, and excludes the velocity range of \hi\ emission from the Magellanic Clouds. Overall, the GASKAP-\hi\ survey provides high-resolution observations (with a synthesized beam of $\theta_{b} =$ 30$^{\prime\prime}$) of the local \hi\ gas towards the MC foreground. Assuming a distance of 250 pc to the local gas, a GASKAP beam corresponds to a linear size of $\sim$0.04 pc, which will allow us to study the physical properties of the gas in high detail and explore its relationship with other interstellar components such as dust. 

\begin{figure*}
 \center
  \includegraphics[width=\textwidth]{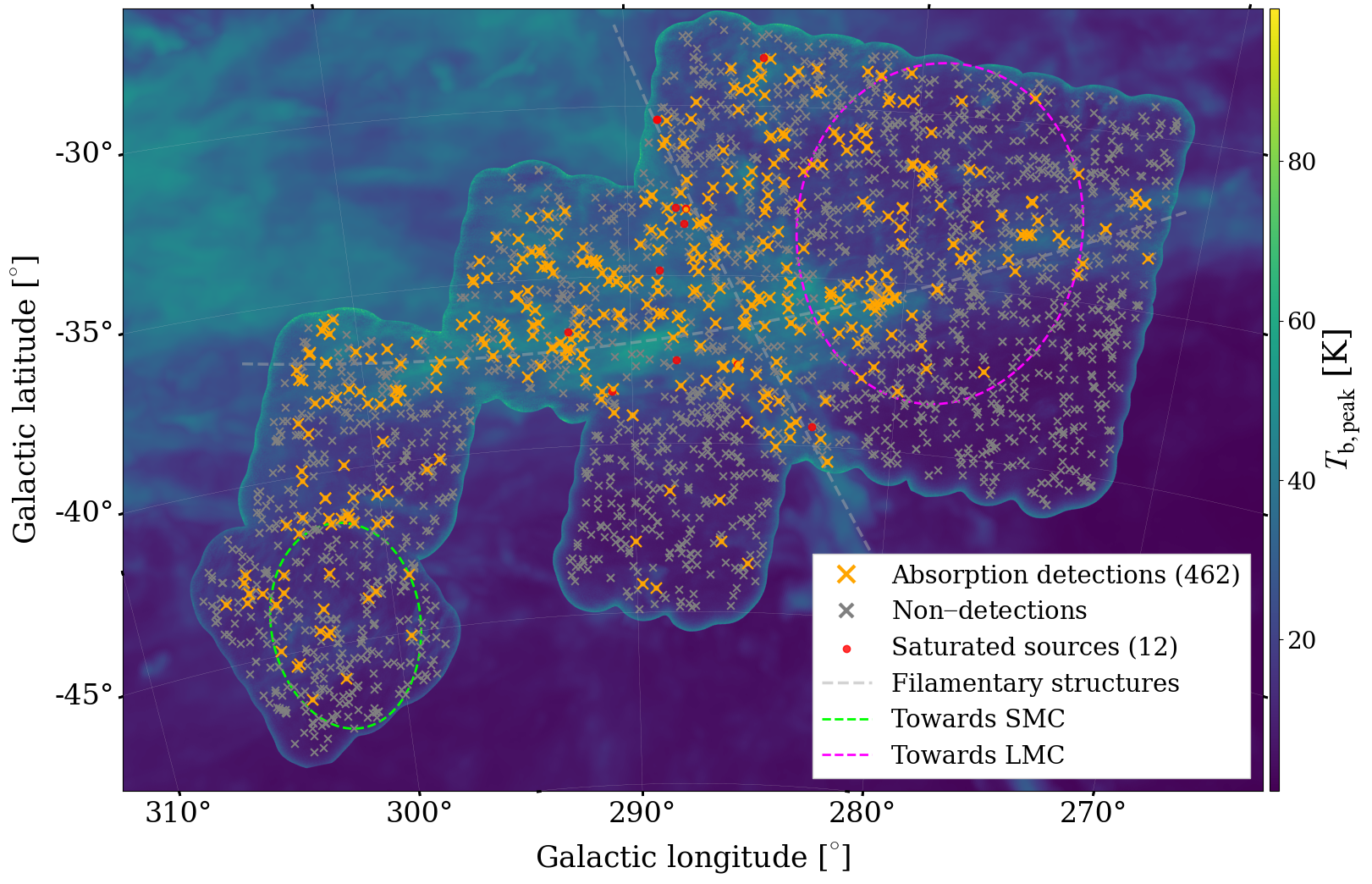}
  \caption{Locations of all \nlos\ background radio continuum sources ($S_\mathrm{1.4\ GHz} \geq$ 15 mJy at 1.4 GHz) in the GASKAP-\hi\ Pilot II survey, overlaid on the GASKAP peak brightness temperature map (\TBpeak, with an angular resolution of 30$^{\prime\prime}$). The light blue outer boundary shows the GASKAP-\hi\ 250-square-degree observing footprint (11 sources per square degree), the GASS peak brightness is shown in the background outside this footprint. Orange crosses mark \ndet\ lines of sight with absorption detections at the 3$\sigma$ threshold, gray crosses indicate non-detections (2239), and red dots show the saturated cases (12). The approximate directions toward the LMC and SMC are represented by magenta and green circles respectively. Two local \hi\ filaments are depicted as gray dashed curves (Reticulum in the vertical direction and Hydrus in the horizontal direction).
  }\vspace{0.4cm}
  \label{fig:all_src_locations}
\end{figure*}

The survey covered an area of approximately 250 deg$^{2}$, spanning a Galactic latitude range of $b$ = ($-45^{\circ}$, $-27^{\circ}$) and a Galactic longitude range of $l$ = ($270^{\circ}$, $305^{\circ}$). The light-blue outer boundary in Figure \ref{fig:all_src_locations} illustrates the GASKAP observing footprint. In this area, \hi\ absorption spectra were extracted towards 2,714 background continuum sources with flux densities above 15 mJy at 1.4 GHz, which leads to a source density of 11 per deg$^{2}$. Of these, 462 sources exhibit absorption features at the 3$\sigma$ detection threshold in the Galactic velocity range. This equates to a 17\% absorption detection rate, or approximately two detections per deg$^{2}$. We excluded 12 sightlines from the analysis due to their saturated opacity spectra, which result in extremely high optical depths with enormous uncertainties.
% See hi03_read_GASKAP_abs_DR3.ipynb, Check saturated l.o.s

% See gfit: hires/hi26_abs_dr3.ipynb
% See gfit: hires/hi23_summmary.ipynb

\subsection{Extracting \hi\ absorption}
\label{subsec:hi_absorption}
We employ the GASKAP-\hi\ absorption pipeline developed by \cite{Dempsey2022} to extract \hi\ absorption spectra for \nlos\ sources. This pipeline takes as inputs a collection of calibrated ASKAP beam measurement sets and employs the ASKAP Selavy source finder \citep{Whiting2012} (which operates on the continuum images) to generate the GASKAP-\hi\ continuum source catalog. The continuum source catalog determines a source ellipse for each component. The majority (98.7\%, 2680 out of 2714) of the sources have a major axis of $<$15$^{\prime\prime}$, with a few having larger major axes of up to 60$^{\prime\prime}$.

For any continuum sources that exceed a continuum flux threshold ($S_\mathrm{cont} \geq$ 15 mJy at 1.4 GHz), we first extract the on-source spectra from all pixels within the source ellipse.
% by combining all pixels within the corresponding source ellipse. 
Next, we define a line-free region in the data cube, specifically within the velocity range of $-100$ to $-60$ \kms\ for the foreground of the MCs. Within this region, we calculate the mean continuum flux density
% brightness
for each pixel. We then generate a weighted-mean on-source spectrum by combining the spectra from all pixels (as described in \cite{Dickey1992}), where the weights are computed as the square of the pixel continuum flux density. Finally, to obtain the absorption spectrum  ($e^{-\tau_{v}}$), we normalize the combined spectrum by dividing it by the mean flux density within its line-free region.

We estimate the noise in the absorption spectrum by considering both the noise in the off-line region and the emission within the primary beam of the dish \citep{Jameson2019,Dempsey2022}. The standard deviation of the spectrum in the line-free region serves as the base noise measure. Additionally, we incorporate the emission data from the Parkes Murriyang GASS survey \citep{McClureGriffiths2009} to account for the increase in system temperature caused by the received emission. Here, we average the GASS emission across a 33$^{\prime}$ radius annulus centered on the source position, excluding the central 1 pixel ($\sim$5$^{\prime}$). We calculate the 1$\sigma$ noise profile for the GASKAP absorption spectrum as:

\begin{equation}
\sigma_\mathrm{\tau} (v) = \sigma_\mathrm{c} \frac{T_\mathrm{sys} + \eta_\mathrm{ant}T_\mathrm{em} (v)}{T_\mathrm{sys}},
\label{eq:abs_noise_spectrum}
\end{equation}
where $T_\mathrm{sys}$ is the system temperature (50 K) and $\eta_\mathrm{ant}$ is the antenna efficiency (0.67) based on \cite{Hotan2021}. Here, $\sigma_\mathrm{c}$ represents the standard deviation of the line-free region of the absorption spectrum, and the mean emission $T_\mathrm{em} (v)$ at each velocity step is obtained from GASS measurements.

The GASKAP absorption data, initially acquired with a native spectral resolution of 0.244 \kms, were subsequently smoothed to a spectral resolution of 0.977 \kms\ (4 times the native bin width).
% to align with the spectral resolution of the GASS survey.
Finally, we analyze all spectra to identify significant absorption detections at the 3$\sigma$ threshold and provide a catalog of spectra with absorption features. %DL clarification
% The average integration time per source using ASKAP wide-field imaging is 10 hours, resulting in a
Our noise level in optical depth varies between 0.002 and 0.33 per 0.977 \kms\ velocity channel, with a mean and median of 0.05 and 0.04, respectively.

% $T_\mathrm{sky}$ 
Figure \ref{fig:all_src_locations} displays the positions of \ndet\ sources with detected absorption in Galactic coordinates, and Table \ref{table:source_list} provides their basic information, including scheduled block ID (SBID), right ascension (R.A.), declination (DEC), Galactic longitude/latitude, flux density at 1.4 GHz, the median root mean square (rms) uncertainty in optical depth per 0.977 km s$^{-1}$ channel, and diffuse background radio continuum emission (\Tbg, the background brightness temperature including the 2.725 K isotropic radiation from the CMB and the Galactic synchrotron background at the source position) for a sample of 50 lines of sight. A full table containing information for all \ndet\ sources is available in the online version of this publication. All GASKAP emission and absorption spectra, along with the fitted quantities (optical depths, CNM spin temperatures, WNM and CNM Gaussian parameters), their associated uncertainties, and the data analysis notebooks are publicly available\footnote{https://github.com/GASKAP/HI-Absorption}.

% See hi25_summary_stats.ipynb
%srclist.py
\begin{table*}%[htbp]
\begin{center}
\fontsize{8}{7}\selectfont
\caption{Basic information of \ndet\ sightlines (full table for all sources is available in the online version)}
\centering
\label{table:source_list}
\begin{tabular}{lccccclll}
\noalign{\smallskip} \hline \hline \noalign{\smallskip}
\shortstack{Source\\ $\ $} & \shortstack{SBID\\(scheduled block)} & \shortstack{R.A (J2000)\\(hh:mm:ss)} & \shortstack{DEC (J2000)\\(dd:mm:ss)} & \shortstack{$l$\\$(^{o})$} & \shortstack{$b$\\$(^{o})$} &
\shortstack{$S_\mathrm{1.4\ GHz}$\\(mJy)} & \shortstack{$\sigma_{\tau}$\\ $\ $} & \shortstack{$T_\mathrm{bg}$\\(K)} \\
\hline

J001424$-$733911 & 30665 & 00:14:24 & $-$73:39:11 & 306.49 & $-$43.22 & 179.5 & 0.015 & 3.56 \\
J002144$-$741500 & 30665 & 00:21:44 & $-$74:14:59 & 305.67 & $-$42.72 & 132.4 & 0.028 & 3.56 \\
J002223$-$742825 & 30665 & 00:22:23 & $-$74:28:26 & 305.56 & $-$42.51 & 206.7 & 0.019 & 3.56 \\
J002248$-$734007 & 30665 & 00:22:48 & $-$73:40:07 & 305.69 & $-$43.31 & 49.8 & 0.058 & 3.56 \\
J002335$-$735529 & 30665 & 00:23:35 & $-$73:55:29 & 305.57 & $-$43.06 & 61.6 & 0.019 & 3.57 \\
J002337$-$735529 & 30665 & 00:23:37 & $-$73:55:30 & 305.56 & $-$43.06 & 128.3 & 0.017 & 3.57 \\
J002907$-$735349 & 30665 & 00:29:07 & $-$73:53:49 & 305.05 & $-$43.14 & 74.9 & 0.028 & 3.57 \\
J003412$-$733314 & 30665 & 00:34:12 & $-$73:33:15 & 304.61 & $-$43.52 & 49.8 & 0.030 & 3.59 \\
J003414$-$733327 & 30665 & 00:34:14 & $-$73:33:28 & 304.61 & $-$43.51 & 292.0 & 0.020 & 3.59 \\
J003424$-$721143 & 30665 & 00:34:25 & $-$72:11:43 & 304.77 & $-$44.87 & 156.8 & 0.019 & 3.55 \\
J003749$-$735127 & 30665 & 00:37:50 & $-$73:51:27 & 304.23 & $-$43.24 & 20.0 & 0.104 & 3.59 \\
J003824$-$742211 & 30665 & 00:38:25 & $-$74:22:11 & 304.13 & $-$42.73 & 227.2 & 0.010 & 3.50 \\
J003939$-$714141 & 30665 & 00:39:40 & $-$71:41:41 & 304.25 & $-$45.40 & 74.9 & 0.025 & 3.56 \\
J004048$-$714600 & 30665 & 00:40:48 & $-$71:45:59 & 304.12 & $-$45.34 & 463.3 & 0.009 & 3.56 \\
J004222$-$754838 & 38215 & 00:42:22 & $-$75:48:38 & 303.67 & $-$41.30 & 763.1 & 0.008 & 3.47 \\
J004222$-$754838 & 30665 & 00:42:23 & $-$75:48:38 & 303.67 & $-$41.30 & 763.1 & 0.008 & 3.47 \\
J004330$-$704147 & 30665 & 00:43:31 & $-$70:41:48 & 303.88 & $-$46.42 & 310.2 & 0.012 & 3.55 \\
J004741$-$753010 & 30665 & 00:47:41 & $-$75:30:11 & 303.25 & $-$41.62 & 335.0 & 0.013 & 3.47 \\
J004957$-$723554 & 30665 & 00:49:57 & $-$72:35:54 & 303.09 & $-$44.53 & 92.0 & 0.036 & 3.77 \\
J005219$-$722705 & 30665 & 00:52:19 & $-$72:27:05 & 302.84 & $-$44.68 & 72.4 & 0.020 & 3.77 \\
J005238$-$731244 & 30665 & 00:52:38 & $-$73:12:44 & 302.81 & $-$43.92 & 107.7 & 0.029 & 3.75 \\
J005321$-$770019 & 38215 & 00:53:22 & $-$77:00:19 & 302.79 & $-$40.12 & 70.7 & 0.055 & 3.43 \\
J005337$-$723143 & 30665 & 00:53:38 & $-$72:31:43 & 302.70 & $-$44.60 & 68.5 & 0.047 & 3.81 \\
J005341$-$771713 & 38215 & 00:53:42 & $-$77:17:13 & 302.77 & $-$39.84 & 231.6 & 0.025 & 3.43 \\
J005611$-$710706 & 30665 & 00:56:11 & $-$71:07:06 & 302.38 & $-$46.00 & 442.5 & 0.005 & 3.54 \\
J005641$-$783945 & 38215 & 00:56:42 & $-$78:39:45 & 302.60 & $-$38.46 & 155.7 & 0.033 & 3.47 \\
J005732$-$741242 & 30665 & 00:57:33 & $-$74:12:43 & 302.37 & $-$42.91 & 461.7 & 0.008 & 3.49 \\
J010120$-$781900 & 38215 & 01:01:21 & $-$78:18:60 & 302.29 & $-$38.80 & 106.5 & 0.026 & 3.47 \\
J010214$-$801239 & 38215 & 01:02:15 & $-$80:12:39 & 302.36 & $-$36.91 & 230.3 & 0.035 & 3.50 \\
J010218$-$754651 & 38215 & 01:02:19 & $-$75:46:51 & 302.04 & $-$41.33 & 275.5 & 0.014 & 3.43 \\
J010218$-$754650 & 30665 & 01:02:19 & $-$75:46:51 & 302.04 & $-$41.33 & 261.2 & 0.011 & 3.43 \\
J010236$-$762313 & 38215 & 01:02:37 & $-$76:23:13 & 302.06 & $-$40.72 & 58.8 & 0.034 & 3.41 \\
J010238$-$762315 & 38215 & 01:02:38 & $-$76:23:16 & 302.06 & $-$40.72 & 74.7 & 0.064 & 3.41 \\
J010238$-$762315 & 30665 & 01:02:38 & $-$76:23:15 & 302.06 & $-$40.72 & 74.7 & 0.064 & 3.41 \\
J010249$-$795604 & 38215 & 01:02:49 & $-$79:56:05 & 302.31 & $-$37.18 & 284.5 & 0.011 & 3.49 \\
J010249$-$795600 & 38215 & 01:02:50 & $-$79:55:59 & 302.31 & $-$37.18 & 86.8 & 0.014 & 3.49 \\
J010251$-$753523 & 38215 & 01:02:51 & $-$75:35:24 & 301.98 & $-$41.52 & 78.8 & 0.054 & 3.44 \\
J010251$-$753523 & 30665 & 01:02:52 & $-$75:35:23 & 301.98 & $-$41.52 & 78.8 & 0.054 & 3.44 \\
J010452$-$795246 & 38215 & 01:04:53 & $-$79:52:47 & 302.19 & $-$37.23 & 35.3 & 0.124 & 3.49 \\
J010811$-$754156 & 30665 & 01:08:12 & $-$75:41:56 & 301.55 & $-$41.38 & 25.3 & 0.105 & 3.46 \\
J010912$-$790840 & 38215 & 01:09:12 & $-$79:08:41 & 301.87 & $-$37.95 & 103.3 & 0.040 & 3.48 \\
J011005$-$722647 & 30665 & 01:10:05 & $-$72:26:48 & 300.96 & $-$44.61 & 144.5 & 0.019 & 3.65 \\
J011049$-$731427 & 30665 & 01:10:50 & $-$73:14:27 & 301.00 & $-$43.81 & 596.7 & 0.005 & 3.57 \\
J011133$-$753809 & 38215 & 01:11:34 & $-$75:38:09 & 301.27 & $-$41.43 & 89.0 & 0.061 & 3.47 \\
J011134$-$753809 & 30665 & 01:11:34 & $-$75:38:09 & 301.27 & $-$41.43 & 90.3 & 0.039 & 3.47 \\
J011136$-$753800 & 30665 & 01:11:36 & $-$75:38:00 & 301.27 & $-$41.43 & 69.9 & 0.045 & 3.47 \\
J011321$-$752819 & 30665 & 01:13:21 & $-$75:28:20 & 301.10 & $-$41.58 & 107.1 & 0.036 & 3.46 \\
J011344$-$803517 & 38215 & 01:13:45 & $-$80:35:17 & 301.80 & $-$36.49 & 371.3 & 0.017 & 3.57 \\
J011348$-$803455 & 38215 & 01:13:48 & $-$80:34:56 & 301.80 & $-$36.50 & 87.2 & 0.059 & 3.57 \\
J011432$-$732142 & 30665 & 01:14:33 & $-$73:21:43 & 300.65 & $-$43.66 & 164.1 & 0.019 & 3.56 \\
J011440$-$803741 & 38215 & 01:14:41 & $-$80:37:42 & 301.76 & $-$36.45 & 117.2 & 0.035 & 3.57 \\
J011549$-$771147 & 38215 & 01:15:50 & $-$77:11:48 & 301.17 & $-$39.85 & 199.3 & 0.022 & 3.43 \\

\noalign{\smallskip} \hline \noalign{\smallskip}

% \multicolumn{7}{l}{\textsuperscript{*}\footnotesize{\citet{Whiting2012}}} \\
% {\textsuperscript{**}\footnotesize{A full table containing information for all 528 sources is available in the online version}}

\end{tabular}
\end{center}
\end{table*}

\subsection{Matching \hi\ emission}
\label{subsec: matching_hi_emission}

In order to obtain the physical properties of the atomic ISM, in addition to the absorption measurements, it is essential to incorporate constraints from the \hi\ emission. Ideally, one should measure the emission of the \hi\ gas in the precise direction of the continuum background source, but this is impossible because the continuum background source cannot be turned off. We must therefore measure the \hi\ emission in the vicinity of the background source and assume that the emission and absorption measurements originate from the same \hi\ gas clouds. This assumption is reasonable for the present observations of \hi\ gas in the Milky Way; with the unprecedented angular resolution of the GASKAP-\hi\ survey, the \hi\ emission can be observed very close ($\lesssim$ 60$^{\prime\prime}$, or linear distance $\lesssim$ 0.07 pc) to the continuum background sources. Since nearly all of the sources (98.7\% as stated above) have size $<$30$^{\prime\prime}$, \hi\ emission observations are acquired roughly on the same angular scale as \hi\ absorption. We also smooth GASKAP emission data to a 0.977 \kms\ spectral resolution to align with that of the GASKAP absorption. Given the high noise ($\sim$1 -- 1.5 K, ) in the GASKAP-\hi\ Pilot II emission data (which accompanies its high spatial resolution as a trade-off), we further resample both the \hi\ absorption and emission spectra into 0.2 \kms\ channel bins using linear interpolation before their Gaussian decomposition.  This resampling does not affect the absorption features, as they are robustly detected with the 3$\sigma$ threshold discussed in Section \ref{subsec:hi_absorption}. However, it facilitates a more accurate and stable determination of the Gaussian emission features during the fitting process.

\begin{figure}
\includegraphics[width=1.0\linewidth]{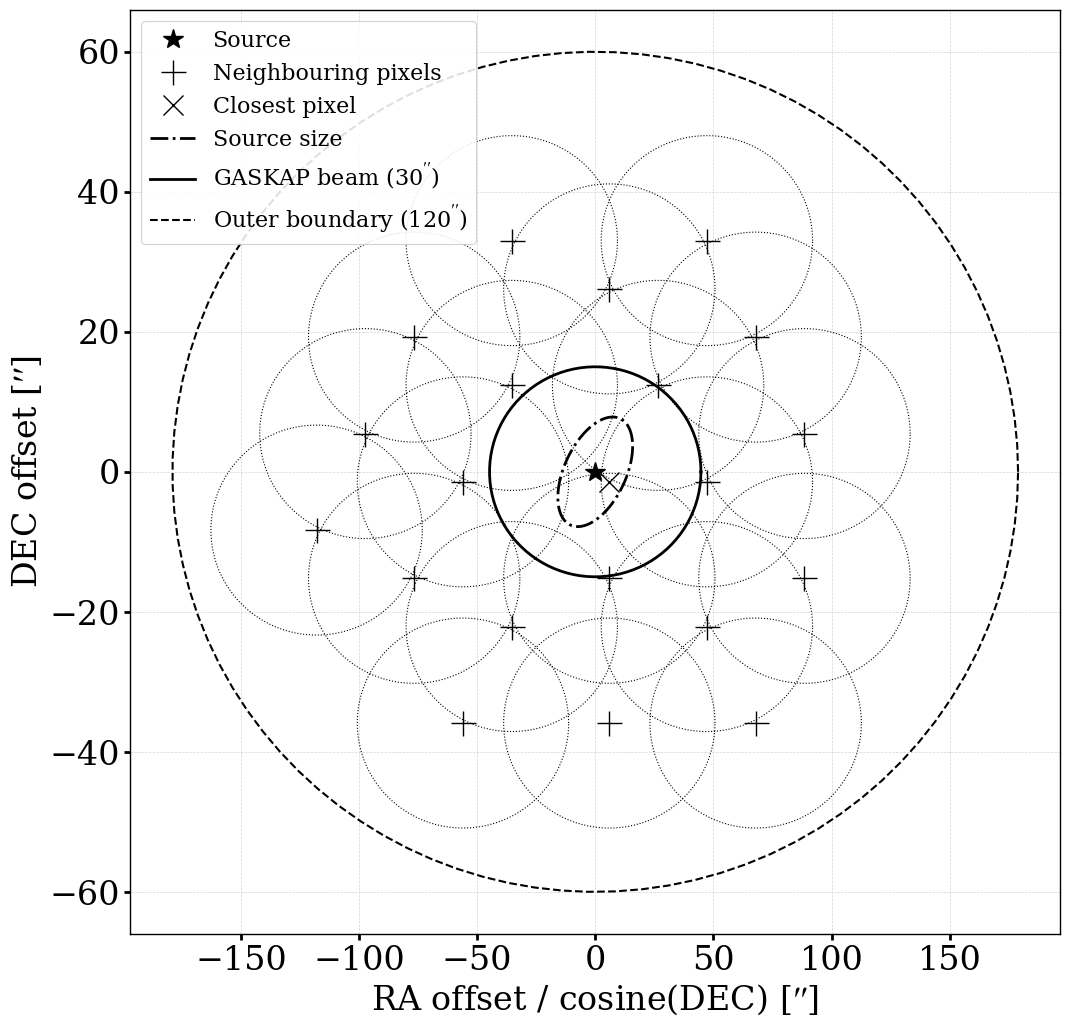}
\caption{Example of the distribution of \hi\ emission measurements around a continuum source (star at the center). The ``\texttt{+}'' marks represent selected neighboring GASKAP emission spectra. The ``X'' indicates the nearest pixel in the GASKAP emission datacube to the continuum source. Only one spectrum per GASKAP synthesized beam (30$^{\prime\prime}$) was chosen to mitigate their correlation within a beam. The solid/dotted circles show the GASKAP beam size, with the dashed circle representing the outer boundary at an angular distance of two beam FWHM (60$^{\prime\prime}$) from the source.}
\label{fig:em_pixels}
\end{figure}

For each target, we first extract emission spectra within a radius of 60$^{\prime\prime}$, equivalent to $\sim$315 pixels in the GASKAP data. In general, this indicates that the average emission is sampled over an effective solid angle of 120$^{\prime\prime}$, which translates to a 0.15 pc linear size (also assuming a distance $d = 250$ pc). We then exclude any spectra within a radius of one GASKAP beam (FWHM) from the continuum source in order to reduce contamination by the source (while still keeping emission spectra as close to the source as possible). Furthermore, to alleviate the correlation of emission measurements within a beam, we retain only one spectrum within each beam size (as illustrated in Figure \ref{fig:em_pixels}). For each source, we retain 20 GASKAP emission spectra at various angular distances to the source. Each emission $T_\mathrm{b} (v)$ spectrum is used as an off-source profile and paired with the absorption profile to estimate (an instance of) the physical properties of \hi\ gas. Ideally, this approach would yield 20 instances of the \hi\ gas properties per source (one for each emission-absorption pair). However, not all emission-absorption pairings successfully converge during fitting, primarily due to the constraints we imposed, such as \Ts\ $>$ $T_\mathrm{bg}$, requiring WNM Gaussian component amplitudes to be above noise levels, and ensuring an opacity correction (ratio between ``true'' column density \NHI\ and optically-thin column density \NHIthin) \RHI\ $\gtrsim$ 1. The distributions of the resulting fitted parameters then naturally incorporate the fluctuations in $T_\mathrm{b}(v)$ at various angular distances around a source, and can be used to derive uncertainties. Finally, the fitted parameters and their uncertainties from successful fits are computed by applying weights inversely proportional to the angular distance between the absorption and emission measurements. 

Along a few lines of sight where the emission spectra are strongly contaminated by absorption features from a strong continuum source, for example, six out of 40 sources in the SBID 33047 centered at R.A = 05h33m00s, DEC = $-$69$^{\circ}$51$^{\prime}$28$^{\prime\prime}$ ($l = 280^{\circ}.4, b = -32^{\circ}.1$), we select the associated emission spectra up to 2$^{\prime}$.5 away from the background sources, and also exclude the inner beams in the annulus.

% See gfit/hi10_extract_em_spectra.ipynb
% Under: Find the neighboring pixels
% Field 8, J035410-702439

\section{Auxiliary data sets}
    \label{subsec:auxiliary_data}
To achieve the project’s objectives, we make use of auxiliary data sets in addition to our primary GASKAP-\hi\ data. These include publicly available \hi\ and dust data. We compare our findings to those from previous Galactic \hi\ absorption studies compiled under the ``BIGHICAT'' meta–catalog by \citet{McClure-Griffiths2023}. We selected \nbiglos\ absorption lines of sight in the Solar neighborhood at high latitudes $|b| > 10^{\circ}$: 40 from 21-SPONGE \citep[][M15 and M18]{Murray2015,Murray2018} focusing on high Galactic latitudes with high optical depth sensitivity, 60 from GMRT04 \citep{Roy2013,Roy2013b}, 16 from MACH \citep{Murray2021}, 51 from Millennium survey (MS) targeting random Galactic lines of sight in the Arecibo sky \citep[][HT03]{Heiles2003a}, 26 from Perseus survey \citep[][S14] {Stanimirovic2014} covering the vicinity of the Perseus molecular cloud, and 71 from GNOMES \citep{Nguyen2019} around the Taurus and Gemini molecular clouds. The dust reddening \ebv\ from \citet{Schlegel1998} (SFD98) and \citet{PLC2014} are utilized to estimate the dust amount along our lines of sight.

\section{\hi\ absorption \& emission Gaussian fitting}
    \label{subsec:gaussian_fitting}

This analysis is based on solving two radiative transfer equations for both on-source and off-source measurements, with the underlying assumption that both on and off positions sample the same gas (as discussed in Section \ref{subsec: matching_hi_emission}):

\begin{equation}
T_\mathrm{b}^\mathrm{ON} (v) = (T_\mathrm{bg}+T_\mathrm{c})e^{-\tau_{v}} + T_\mathrm{s}(1-e^{-\tau_{v}}),
\label{eq:on_simp}
\end{equation}
\begin{equation}
T_\mathrm{b}^\mathrm{OFF} (v) = T_\mathrm{bg}e^{-\tau_{v}} + T_\mathrm{s}(1-e^{-\tau_{v}}),
\label{eq:off_simp}
\end{equation}

\noindent where, $T_\mathrm{b}^\mathrm{ON} (v)$ and $T_\mathrm{b}^\mathrm{OFF} (v)$ are the brightness temperatures of the on-source and off-source profiles, respectively, \Ts\ is the spin temperature, $\tau_{v}$ is the optical depth, and \Tc\ is the brightness temperature of the continuum source.
% and \Tbg\ is the background brightness temperature including the 2.725 K isotropic radiation from the CMB and the Galactic synchrotron background at the source position.
We obtain the background brightness temperature \Tbg\ from the 1.4 GHz radio continuum maps of the Parkes Murriyang CHIPASS survey \citep[angular resolution of 14$^{\prime}$.4,][]{Calabretta2014}, with typical values of around 3.5 K toward the MC foreground. Since the effective distance from a continuum source to the emission spectra is $\sim$1$^{\prime}$ -- significantly smaller than the 14$^{\prime}$.4 CHIPASS resolution, it is reasonable to assume that \Tbg\ is smooth on the scale of the emission-absorption offsets.

In this study, we utilize the HT03 methodology from the Millennium survey to estimate the physical parameters of individual \hi\ clouds along a line of sight. The approach relies on the two-phase neutral atomic medium concept where the observed \hi\ gas consists of two components: cold neutral medium and warm neutral medium. The CNM provides opacity and brightness temperature, whereas the WNM provides only brightness temperature. As a result, both CNM and WNM contribute to the spectrum $T_\mathrm{b} (v)$:
\begin{equation}
 T_\mathrm{b} (v) = T_\mathrm{b, CNM} (v) + T_\mathrm{b, WNM} (v).
\label{eqtb} 
\end{equation}

\noindent Following the HT03 fitting process, we first fit the single on-source absorption spectrum to obtain the Gaussian components of the optical depth. We then perform separate fits for each of the 20 off-source emission spectra (see Figure \ref{fig:em_pixels}), while keeping the optical depth Gaussian components unchanged.

Following \cite{Dempsey2022}, we identify absorption features in the opacity spectra based on specific criteria. An absorption feature is identified if it contains at least one channel with a significance of 3$\sigma$ and an adjacent channel with a significance greater than 2.8$\sigma$. To detect even fainter absorption while reducing the detection of noise spikes, we expand the feature to include any adjacent neighboring channels with a significance of at least 2.8$\sigma$. We then fit the opacity spectrum $\tau(v)$ with a set of $N$ Gaussian components using a least-squares method to determine $\tau_{0,n}$, $v_{0,n}$, and $\delta v_{n}$, which respectively represent the peak optical depth, central velocity, and $\Delta V_\mathrm{FWHM}$ of the n$^{\rm th}$ component.

\begin{equation}
\tau(v) = \sum\limits_{n=0}^{N-1} \tau_{0,n}e^{-4ln2\ [(v-v_{0,n})/\delta v_{n}]^2}.
\label{eqtau} 
\end{equation}

\noindent We assume that each component is independent and isothermal \citep[][]{Wolfire1995}, with a spin temperature $T_\mathrm{s,n}$. The initial guesses for the fit parameters and the number of components are generated from the absorption features identified as described above.

In order to fit each emission profile $T_\mathrm{b}(v)$ (out of 20 emission spectra), we now fix the amplitudes $\tau_0$ derived for the CNM components. We assume that $T_\mathrm{b}(v)$ consists of \textit{N} cold components seen in the absorption spectrum, plus any warm components seen only in emission. The least squares technique is used to fit $T_\mathrm{b}(v)$, allowing the central velocity and width of the CNM components ($v_{0,n}$ and $\delta v_{n}$) to vary within a limited range. The contribution of the cold components is given by,

\begin{equation}
T_\mathrm{b,CNM}(v) = \sum\limits_{n=0}^{N-1} T_{s,n}(1-e^{-\tau_{n}(v)}) e^{-\sum\limits_{m=0}^{M-1} \tau_{m}(v)}
\label{eqtbc} 
\end{equation}
where the subscript $m$ describes $M$ absorption clouds lying in front of the $n^\mathrm{th}$ cloud.

For WNM, we use a set of $K$ Gaussian functions to represent the contribution to $T_\mathrm{b}(v)$. For each $k^\mathrm{th}$ component, we assume that a fraction $F_\mathrm{k}$ of WNM lies in front of all CNM components, while the remaining ($1-F_{k}$) lies behind and is absorbed by CNM components. Therefore, the brightness temperature from these K WNM components is given by:

\begin{equation}
\begin{split}
T_\mathrm{b,WNM}(v) =\ & \sum\limits_{k=0}^{K-1} [F_\mathrm{k} + (1-F_\mathrm{k})e^{-\tau_{v}}]\\ & \times T_\mathrm{0,k}e^{-4ln2\ [(v-v_\mathrm{0,k})/\delta v_\mathrm{k}]^2}
\label{eqtbw}
\end{split}
\end{equation}
where $T_\mathrm{0,k}$, $v_\mathrm{0,k}$, and $\delta v_\mathrm{k}$ are respectively the peak brightness temperature, central velocity and FWHM of the $k^\mathrm{th}$ emission component. This least-squares fit to the emission spectrum gives $T_\mathrm{0,k}$, $v_\mathrm{0,k}$, $F_\mathrm{k}$ for warm components and $T_\mathrm{s,n}$ for cold components.

\begin{figure}
\centering
\includegraphics[width=1.\linewidth]{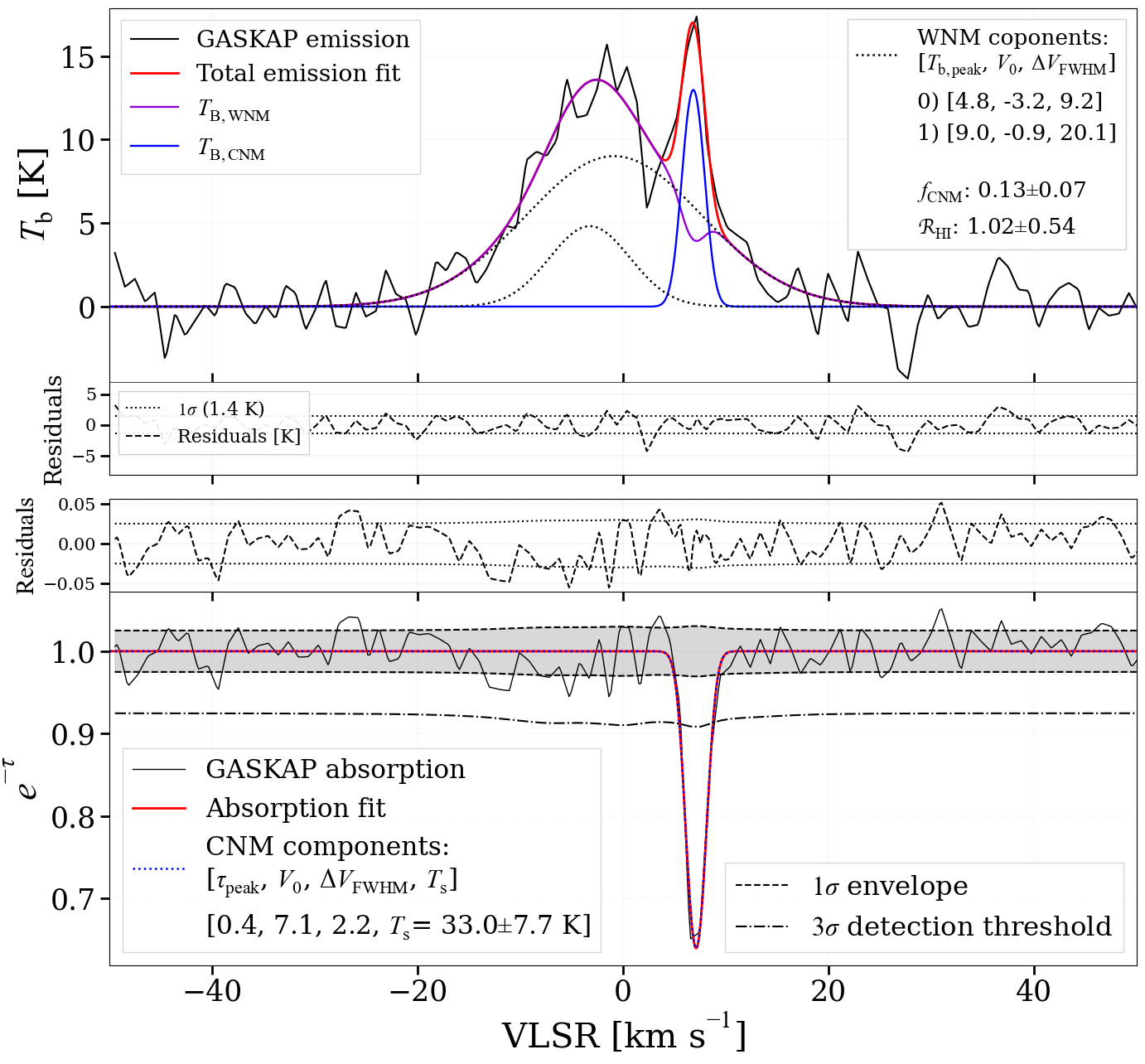}
\caption{Example of Gaussian fits to a pair of emission and absorption spectra (source J054427--715528 in SBID 33047 centered at $l = 280^{\circ}.4, b = -32^{\circ}.1$). In the upper panels, the black solid line represents the emission profile $T_\mathrm{b}(v)$; the black dotted lines indicate the WNM Gaussian components (amplitude in K, central velocity and $\Delta V_\mathrm{FWHM}$ in \kms); the magenta line shows the contribution of the WNM components to the emission profile; the blue line displays the contribution to the emission profile by the CNM components; the solid red line depicts the total $T_\mathrm{b}$ fit. The CNM fraction (\FCNM) and opacity correction factor (\RHI) are also reported. The residuals from the fit are plotted below the $T_\mathrm{b}(v)$ curve, with $\pm\sigma_{T_\mathrm{b}}$ overlaid. In the lower panels, the black line shows the optical depth profile (e$^{-\tau_{v}}$); the dotted lines correspond to the CNM Gaussian components (amplitude \taupeak, central velocity and $\Delta V_\mathrm{FWHM}$ in \kms, \Ts\ in K), and the red line represents the fit to the e$^{-\tau_{v}}$ profile. The dashed lines represent the $1\sigma_{\mathrm{e}^{-\tau}}$, and the dashed--dotted line shows the $3\sigma_{\mathrm{e}^{-\tau}}$ detection limit. The residuals from the absorption fit are presented above the optical depth curve, with $\pm\sigma_{\mathrm{e}^{-\tau}}$ superimposed. The derivation of uncertainty profiles ($\sigma_{\mathrm{e}^{-\tau}}$) is described in Section \ref{subsec:hi_absorption}.}
\label{fig:eg_plot1}
\end{figure}

For each sightline and each emission spectrum (out of 20 per sightline), we perform the $T_\mathrm{b}(v)$ fit with all possible orderings of the $N$ absorption components ($N!$), and choose the one that gives the smallest residuals. The fraction $F_\mathrm{k}$ determines the contribution of $T_\mathrm{b,WNM}(v)$ to the brightness $T_\mathrm{b}(v)$, and thus has an important effect on the derived CNM spin temperatures ($T_\mathrm{s,n}$). After including $K$ WNM emission-only components, we calculate the spin temperature of CNM by assigning three values (0, 0.5, 1) to each $F_\mathrm{k}$. We then experiment with all possible combinations of $K$ WNM components ($3^K$) along the line of sight. We obtain the final spin temperatures $T_\mathrm{s,n}$ for CNM components as a weighted average over all trials, with the weight of each trial being the inverse of the variance estimated from the residuals of the $T_\mathrm{b}(v)$ fit. The uncertainty of the final spin temperature for each CNM component is estimated from the variations in $T_\mathrm{s}$ with $F_\mathrm{k}$ in all trials.
Examples of representative spectra and fits are shown in Figures \ref{fig:eg_plot1}.

This fitting process is repeated for all emission spectra surrounding a continuum source. The grand final fit parameters, along with their uncertainties, are calculated as a weighted average of all fittings, with the weights determined as the inverse of the angular distances of the emission spectra from their associated central sources.

% See gfit/hi20_read_GaussFit_results.ipynb, J054427--715528 in field 33047, use plot_gfit() function

% See hi23_summary_optical_depth.ipynb
\begin{figure}
\includegraphics[width=1.0\linewidth]{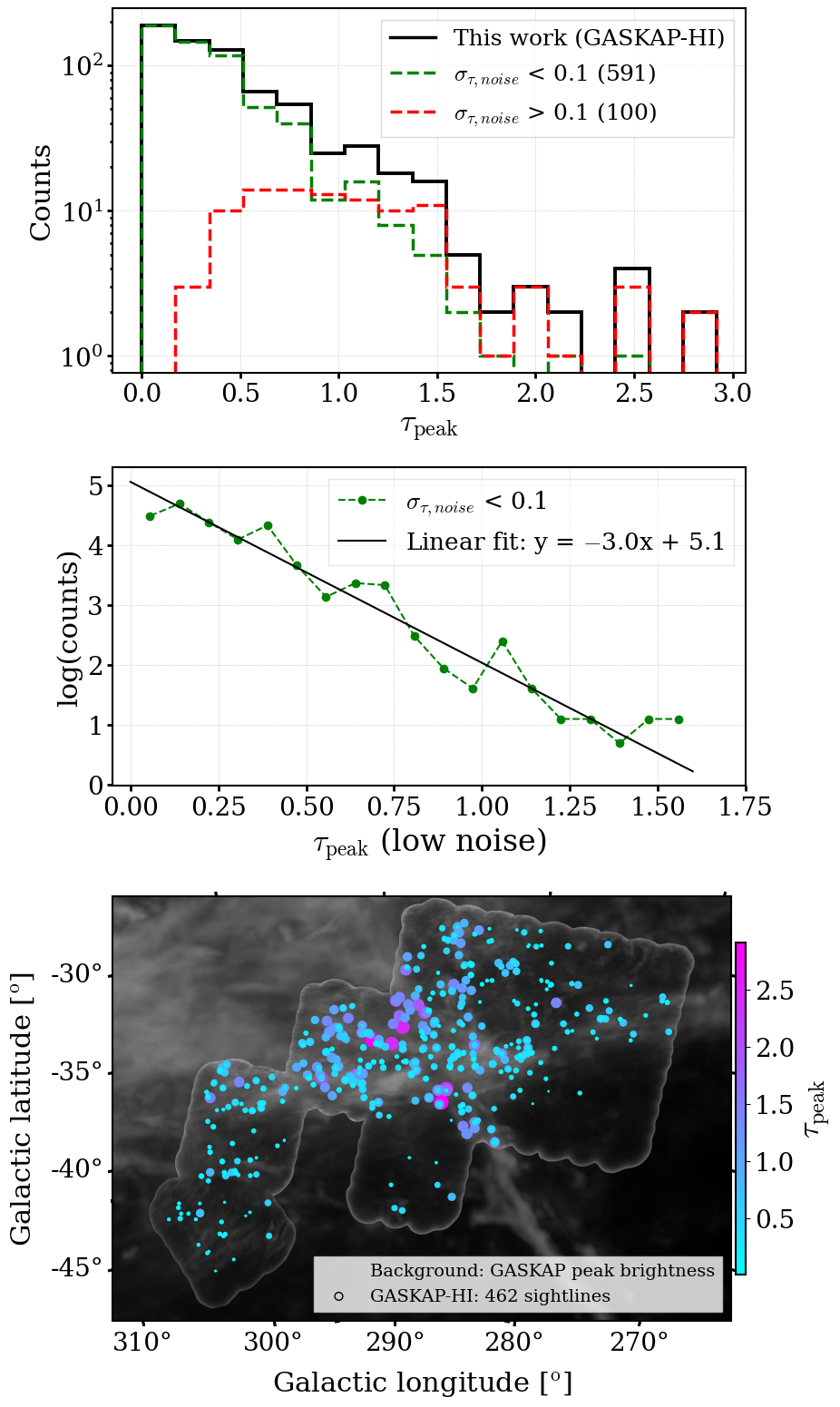}
\caption{Top: \taupeak\ distribution for CNM components in this work (solid line, for \ndet\ lines--of--sight, \ncnm\ CNM components); dashed green line for a subset with low optical depth noise ($\sigma_{\tau} < 0.1$); dashed red line for a subset with higher optical depth noise ($\sigma_{\tau} > 0.1$). Middle: \taupeak\ histogram for the low--noise subset ($\sigma_{\tau} < 0.1$) with natural log scale in ordinate axis; a linear fit to the data is shown by the solid black line. Bottom: The locations of \taupeak\ on the GASKAP--\hi\ peak brightness temperature background (gray); both color and size represent the \hi\ optical depth magnitudes. We define here the ``CNM'' as all \hi\ observed in absorption.}
\label{fig:tau_hist}
\end{figure}

\begin{figure}
\includegraphics[width=1.0\linewidth]{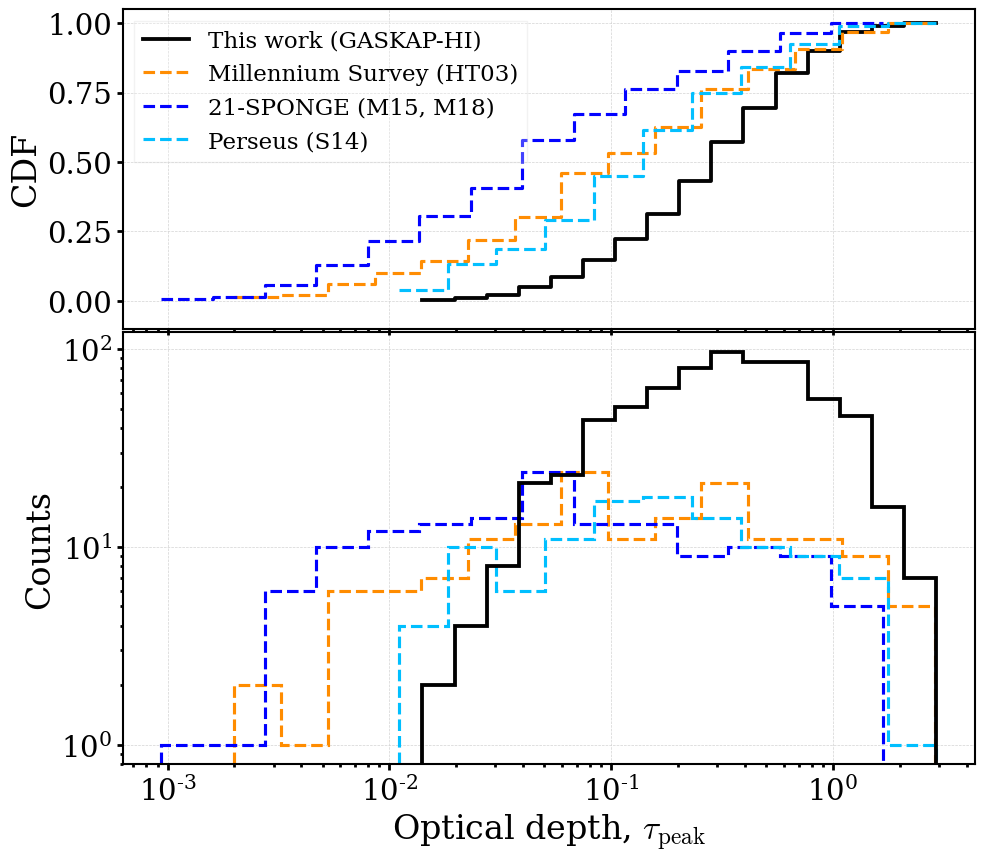}
\caption{Histograms and their cumulative distribution functions (CDF) of CNM peak optical depth \taupeak\ along \ndet\ GASKAP--\hi\ Pilot II sightlines (solid lines). The \taupeak\ temperatures from a subset of \ntsbiglos\ high--latitude BIGHICAT sightlines ($|b| > 10^\circ$) are shown in the dashed line for comparison: 51 from Millennium survey (HT03, orange, 153 CNM components), 40 from 21--SPONGE (M15, blue, 140 CNM components) and 26 from Perseus (S14, cyan, 107 CNM components). The ``CNM'' in all surveys is defined as all \hi\ observed in absorption.}
\label{fig:tau_hists}
\end{figure}

\section{Properties of \hi\ CNM and WNM components}
    \label{results}
    \label{sec:component_properties}
Across the \ndet\ sightlines with \hi\ absorption detections towards the MC foreground, we identify \ncnm\ CNM and $\sim$570 WNM components from the Gaussian fits. Note that the WNM component count is an average value, as we extracted 20 emission spectra around each source and performed the fit with different WNM Gaussian sets for each emission spectrum. The average cold and warm gas fractions derived from the fit results are around 30\% and 70\%, respectively (see Sections \ref{subsec:WNM_temperatures} and \ref{subsec:fcnm} for more details).
% See: hi27_summary_temperatures.ipynb, under section: Summary: read table of integrated properties

In general, we detect absorption and emission in a tight $V_\mathrm{LSR}$ velocity range from $-$10 to 10 \kms\ for absorption and a wider range from $-$30 to 30 \kms\ for emission, with more complex profile shapes found closer to the \hi\ filaments.

Throughout the following sections, we compare our results to previous studies (HT03, S14, M15, and M18). We selected \ntsbiglos\ absorption lines of sight from the literature, corresponding to \ntsbigcnm\ CNM components, in the Solar neighborhood with Galactic latitudes $|b| > 10^{\circ}$. These sightlines probe a wide range of local environmental conditions.

% See hi23_summary_optical_depth.ipynb
\subsection{Optical depth}
\label{subsec: optical_depth}
We derived the peak optical depth, \taupeak, from the Gaussian fitting of the GASKAP-\hi\ absorption spectra. Our \taupeak\ values vary from $\sim$0.014 to 2.9, with a mean of $(0.47\pm0.02)$ and a median of ($0.35\pm0.02$). The uncertainties in the mean and median are obtained by bootstrap resampling. Figure \ref{fig:tau_hist} summarises the \taupeak\ distribution of the full sample of \ncnm\ CNM components (solid line). The green dashed line represents the lower noise (with a standard deviation of $\tau(\mathrm{v})$ noise $\sigma_{\tau} < 0.1$) subsample of 591 CNM components, whereas the red dashed line represents the higher noise ($\sigma_{\tau} > 0.1$) subsample of 100 CNM components. The optical depth distribution peaks at roughly 0.2--0.3, with most of the \taupeak\ (88\%) found below unity. This indicates that the majority of the \hi\ gas in the MC foreground is relatively diffuse. The remaining 12\% of the values, extending in the tail with higher optical depths (1--3), suggest the presence of possible denser gas structures, especially close to the vertical/horizontal \hi\ filaments, as illustrated in the bottom panel. These values in the tail are, however, associated with higher absorption noise and significant uncertainties.

The middle panel of Figure \ref{fig:tau_hist} depicts the distribution of the lower noise subset ($\sigma_{\tau} < 0.1$), where the natural log-scaled frequency is plotted against the \taupeak\ values. The best linear fit to the data is shown by the solid black line. The plot reveals a log-linear pattern in the frequency distribution, indicating an exponential trend in linear space. The bottom panel shows the spatial distribution of the fitted \taupeak, highlighting a concentration of higher optical depth towards gas filaments. Overall, the results indicate a diffuse ISM region in the MC foreground.

Lower noise absorption measurements are more capable of identifying lower optical depths. Figure \ref{fig:tau_hists} illustrates this by comparing the GASKAP-\hi\ Pilot II optical depth with those from four previous absorption studies (as listed in Section \ref{subsec:auxiliary_data}). These studies surveyed various local environments at high Galactic latitudes, $|b| > 10^{\circ}$, using higher opacity sensitivity observations. The \taupeak\ histograms are presented in the lower panel, and their cumulative distribution functions (CDFs) are presented in the upper panel. The GASKAP-\hi\ \taupeak\ distribution is depicted by a solid black line, the distributions from previous surveys are shown by dashed lines.

The GASKAP-\hi\ Pilot II survey exhibits a narrower range of optical depths relative to previous surveys, ranging from 0.014 to 2.9, and does not extend to lower opacity values. This is likely because the previous surveys, with their sensitivities an order of magnitude higher, were able to probe a broader range that includes lower optical depths. The S14 and HT03 surveys, both with relatively similar 1$\sigma$ sensitivities of 10$^{-3}$ per 1 \kms\ velocity channel, seem to show consistent CDFs (no significant difference between their \taupeak\ distributions, K-S test statistic of 0.14, $p$-value of 0.17).
Yet, the S14 Perseus observations missed the lower \taupeak\ part present in the HT03 distribution. S14 attributed this gap partially to their small survey area compared to the HT03 Millennium survey which included more targets at high Galactic latitudes. Additionally, the optical depths around the Perseus molecular cloud do not fall below 10$^{-2}$. Apart from HT03 and S14, all other survey pairs exhibit significant differences in their measured optical depth distributions. In particular, unparalleled sensitivity allowed the 21-SPONGE survey to detect \hi\ optical depths as low as 10$^{-3}$ along high Galactic latitude sightlines, resulting in the broadest distribution of optical depths (Figure \ref{fig:tau_hists}), and the lowest \hi\ opacities (and highest spin temperatures) in all existing absorption surveys (see below).

\subsection{Temperatures of cold and warm gas}
\label{subsec:temperatures}
The spin temperature \Ts\ and kinetic temperature \Tk\ (together with column density) are the primary physical parameters that characterize the thermodynamic properties of neutral atomic ISM. Although \Tk\ is not generally measured, the CNM spin temperature is typically measured to be around 50--100 K. Theoretically, \Ts\ in the CNM is expected to be similar to \Tk\ \citep{Liszt2001}. In contrast, the \hi\ spin temperature is generally expected to be lower than its kinetic temperature in the low-density warm phase, as collisions are insufficient to thermalize the 21-cm transition \citep[][]{Wouthuysen1952,Kulkarni1988,Liszt2001}. Coupling with the background Lyman-$\alpha$ radiation field from Galactic and extragalactic sources can tie \hi\ spin temperature with local gas motions through the Wouthuysen-Field effect \citep[][]{Wouthuysen1952,Field1958PIRE}. Spin temperatures \Ts\ may hence be different from the kinetic temperature depending on the radiation field. But even in the WNM, it is a rough indicator of kinetic temperature.

In the current study, the Gaussian fitting of \hi\ absorption spectra reveals a clear separation between the CNM and WNM components. The $\Delta V_\mathrm{FWHM}$ widths of WNM components are noticeably broader than that of CNM components, with a mean value of 15.0$\pm0.3$ \kms\ for WNM versus 3.1$\pm$0.1 \kms for CNM, respectively.

% Optical depth sensitivity vs Tspin_mean: See gfit/hi26_abs_dr3.ipynb 
\subsubsection{Temperatures of cold gas}
\label{subsec:CNM_temperatures}
\begin{figure}
\includegraphics[width=1.0\linewidth]{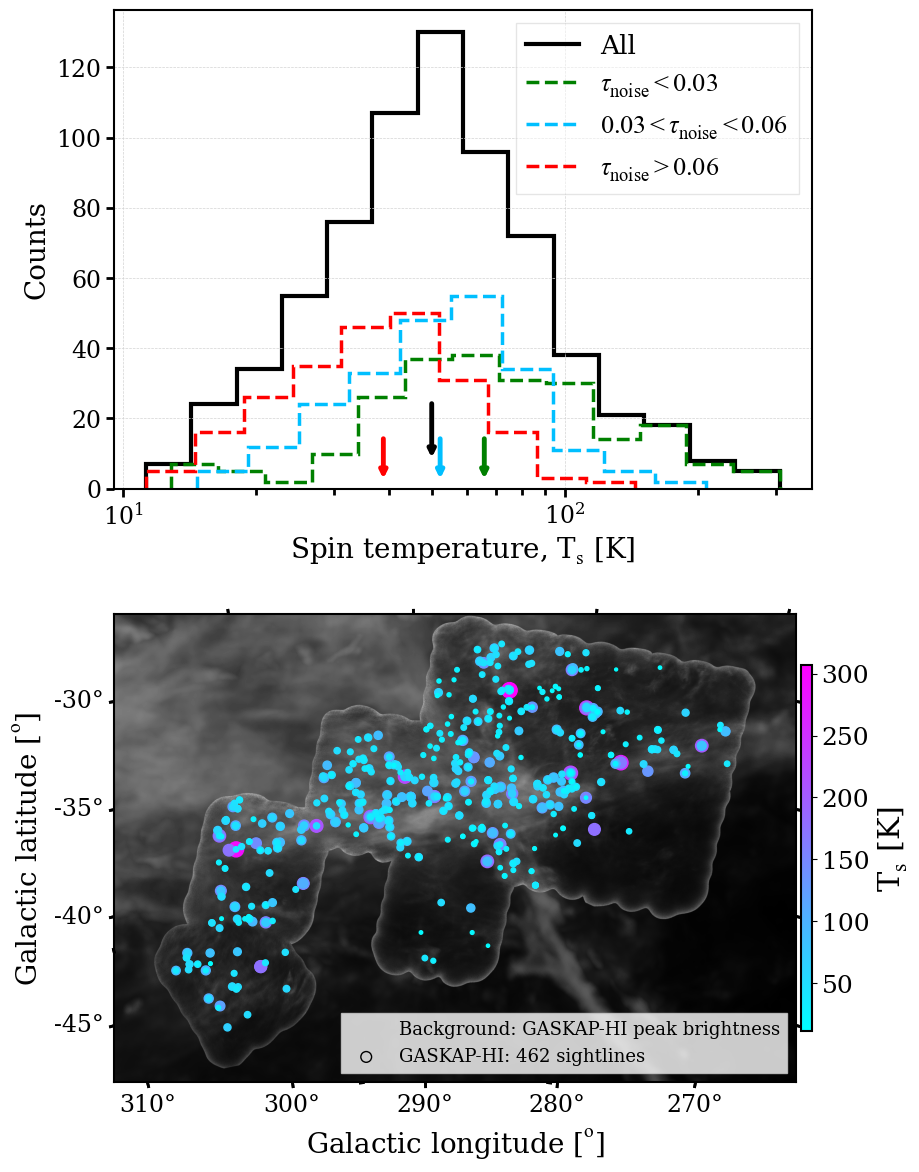}
\caption{Top: Spin temperature histograms for \ncnm\ CNM components along \ndet\ GASKAP-\hi\ sightlines (black solid line), for subsets of high sensitivity (red), intermediate sensitivity (cyan), and low sensitivity (blue). The arrows indicate their associated median spin temperatures. The ``CNM'' in all surveys is defined as all \hi\ observed in absorption. Bottom: Locations of \taupeak\ on the GASKAP-\hi\ peak brightness temperature background (gray, see Figure \ref{fig:all_src_locations} for details); both color and size represent the \hi\ spin temperature magnitudes.}
\label{fig:ts_hist_map}
\end{figure}

% See hi27_summary_temperatures.ipynb
We report the CNM spin temperatures obtained from Gaussian decompositions for our \ndet\ sightlines in Figures \ref{fig:ts_hist_map}. The upper panel displays the CNM \Ts\ distributions for the entire sample (black line) and for three same sample size subsets based on optical depth sensitivities ($\tau_\mathrm{noise}$). In the lower panel, the locations and magnitudes of \Ts\ values are represented on the GASKAP-\hi\ peak brightness map (area within the GASKAP-\hi\ observing footprint).

Our current optical depth sensitivity allows us to accurately measure the temperatures only in the CNM regime, which is comparable with the theoretical predictions of CNM temperatures in a two-phase medium model for Solar neighborhood conditions \citep[20 -- 350 K;][]{Wolfire2003}.
Indeed, we see no obvious tail at \Ts\ $>$ 300 K. The spin temperatures of our \ncnm\ CNM components fall in the range $\sim$11--307 K, with the \Ts\ distribution spreading mostly between 25 K and 130 K, peaking at $\sim$50 K. The mean temperature is 60.3$\pm$1.6 K, the median is at 50.0$\pm$1.2 K, and the column-density-weighted median is 54.8$\pm$1.7 K (uncertainties of the mean/median values are obtained from bootstrap resampling).

\begin{figure*}
\includegraphics[width=0.99\linewidth]{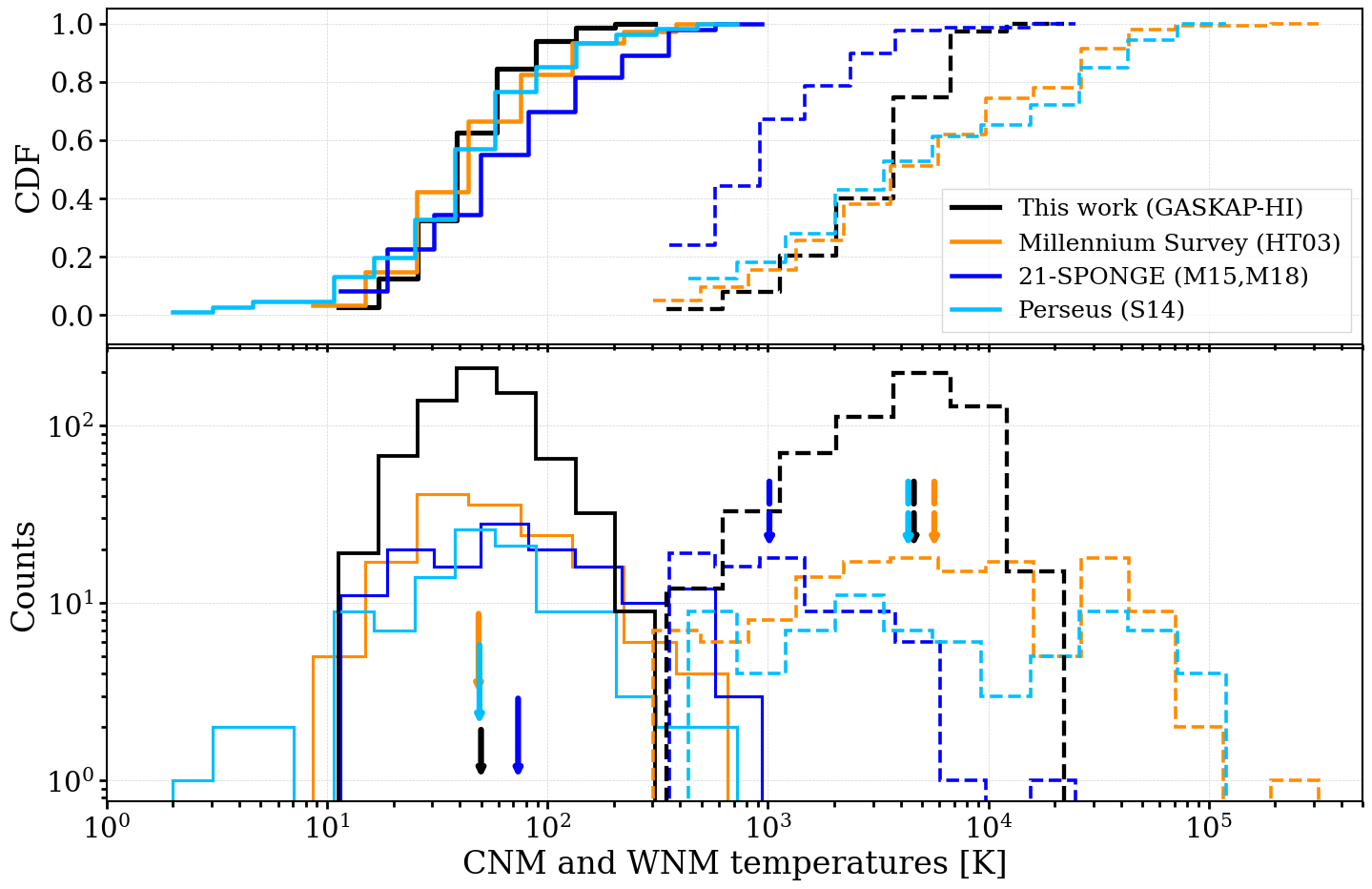}
\caption{Histograms and their CDFs for CNM (\Ts, solid) and WNM (\TD\ or $T_\mathrm{k,max}$, dashed) temperatures along \ndet\ GASKAP-\hi\ sightlines (black). The CNM and WNM temperatures from a subset of \ntsbiglos\ high-latitude BIGHICAT sightlines ($|b| > 10^\circ$) are shown for comparison: 51 from Millennium survey (HT03, orange, 153 CNM components), 40 from 21-SPONGE (M15, blue, 140 CNM components) and 26 from Perseus (S14, cyan, 107 CNM components). The arrows indicate their associated medians of spin (solid) and Doppler (dashed) temperatures. The ``CNM'' in all surveys is defined as all \hi\ observed in absorption, meanwhile the ``WNM'' is identified in emission.}
\label{fig:ts_td_hist}
\end{figure*}

The optical depth sensitivities in the GASKAP-\hi\ absorption survey are not uniform but vary from sightline to sightline (as indicated in Section \ref{sec:observations}), primarily depending on the background source strength. This variation affects the detection of absorption features and ultimately influences the range of derived CNM temperatures. To investigate how the median spin temperature \Ts\ varies with the sensitivity in optical depth, we group CNM spin temperatures into three subsets based on optical depth sensitivity. These subsets have similar sample sizes, and their distributions are also illustrated in the upper panel of Figure \ref{fig:ts_hist_map}: red line for high sensitivity $\sigma_\mathrm{\tau} < 0.03$,  cyan for intermediate sensitivity $0.03 < \sigma_\mathrm{\tau} < 0.06$, and blue for low optical depth sensitivity $\sigma_\mathrm{\tau} > 0.06$. The medians for individual subsets are indicated by corresponding arrows. From high to low sensitivities, their corresponding \Ts\ histograms become narrower accordingly, with spin temperature medians of 65.4$\pm$3.9 K, 52.2$\pm$1.9 K, and 38.7$\pm$1.3 K respectively; the highest \Ts\ values (200 -- 300 K) belong to the highest sensitivity range. The uncertainties of the medians are computed using bootstrap resampling. We observe larger uncertainties for the higher optical depth sample, likely due to their broader \Ts\ distributions. This suggests that the median \Ts\ of CNM features detected in absorption is driven by optical depth sensitivity. We will elaborate on this point below by comparison with the spin temperatures of previous Galactic absorption studies.

Illustrated in Figure \ref{fig:ts_td_hist} is a comparison of CNM spin temperatures (solid lines) obtained from different surveys, but which used similar Gaussian decomposition methodology. In all surveys, ``CNM'' refers to \hi\ observed in absorption. Similar to the optical depth comparison in Section \ref{subsec: optical_depth}, we use the same samples from three previous absorption surveys that probe a wide range of environmental conditions: 140 CNM components from 21-SPONGE (median \Ts\ = 73.6 K); 153 CNM components from MS (median \Ts\ = 48.7 K); and 107 CNM components from the Perseus survey (median \Ts\ = 49.2 K). The lower panel displays histograms with arrows indicating the median \Ts, while the upper panel shows their CDFs.

The GASKAP-\hi\ spin temperatures cover a narrower range compared to other surveys, and are missing higher values beyond 300 K. The previous absorption observations derived \hi\ gas temperatures above 300 K (up to $\sim$700 -- 900 K) in the UNM regime, leading to broader \Ts\ distributions. S14 shows a handful of CNM components with \Ts\ $< 10$ K near the Perseus molecular cloud where the \hi\ temperatures are as low as $\sim$10 K \citep[e.g.,][]{Tielens2005}. Several studies, such as \citealt{Knapp1972}, HT03, \citealt{Meyer2006, Meyer2012} and M15,  also observed these \Ts\ values toward either molecular clouds or low-latitude lines of sight. While these extremely cold CNM components may be due to spurious fits, they could also arise when photoelectric heating by dust grains is absent \citep[e.g.,][]{Wolfire1995,Heiles2003b}. Similar to the comparison of peak optical depths (see Figure \ref{fig:tau_hists}), only the spin temperature CDFs from the S14 and HT03 surveys are likely drawn from the same distribution (K-S test statistic $\approx$ 0.1, $p-$value $\approx$ 0.4). Nevertheless, apart from the higher \Ts\ tails observed in previous works, \Ts\ histograms from all four surveys appear generally similar.

The median spin temperature of the GASKAP-\hi\ data (50.0$\pm$1.2 K), is in agreement with HT03 (48.7$\pm$4.3 K) and S14 (49.2$\pm$3.9 K), as seen in Figure \ref{fig:ts_td_hist}. It is also consistent with the GNOMES survey (54.6$\pm$2.1 K) in the vicinity of the Taurus and Gemini molecular clouds \citep{Nguyen2019}, and \cite{Denes2018} (48.0 K) in the Riegel-Crutcher cloud. However, our measured \Ts\ medians are lower than those reported in the 21--SPONGE survey (M15, M18), which are 73.6$\pm$8.7 K for their entire CNM sample, mainly due to their improved sensitivity to optical depth.

These patterns suggest that the spin temperature distribution of the CNM remains relatively consistent throughout the local ISM under the heating/cooling equilibrium for Solar neighborhood conditions. Any discrepancies in the \Ts\ distributions could mainly be attributed to variations in sensitivity in the absorption measurements: Higher sensitivity observations can detect higher \Ts\ in absorption (as highlighted by M15 and M18).

The ongoing GASKAP-\hi\ main survey in full mode, with an integration time 20 times longer than the current Pilot II observation and achieving a sensitivity $\sim$4.5 times higher, is likely to measure higher spin temperatures, potentially falling within the range of the unstable neutral medium. Therefore, upcoming GASKAP-\hi\ studies will provide an opportunity to investigate the transition from UNM to CNM in the MC foreground region.

\subsubsection{Temperatures of warm gas}
\label{subsec:WNM_temperatures}
For the warm gas components, we calculate the Doppler temperature from their line widths using the formula introduced by \cite{Payne1982}, $T_\mathrm{D} = 21.86 \times \Delta V^2_\mathrm{FWHM}$ K. Since the observed $\Delta V_\mathrm{FWHM}$ primarily results from a combination of thermal and turbulent broadenings, the Doppler temperature indicates an upper limit on the gas kinetic temperature \Tkmax. Dashed black lines in Figure \ref{fig:ts_td_hist} depict the GASKAP-\hi\ \TD\ histogram and its CDF for WNM components in log-log space. The mean and median (black dashed arrow) are $\sim$4900 K ($\Delta V_\mathrm{FWHM} \simeq 15$ \kms) and $\sim$4600 K ($\Delta V_\mathrm{FWHM} \simeq 14.5$ \kms), respectively. All WNM lies between $\sim$365 K and $\sim$20,000 K ($\Delta V_\mathrm{FWHM} \sim$ 4 \kms\ to 30 \kms), and the \TD\ distribution shows no clear tail above 10,000 K. With a negligible gap between CNM \Ts\ and WNM \TD\ from 307 K to 365 K, the GASKAP-\hi\ CNM and WNM temperatures form a continuous distribution showing two distinct populations for cold and warm \hi, peaking respectively at $\sim$50 K and $\sim$4600 K. Between the two peaks is a valley of intermediate temperatures ($\sim$250 K -- 4000 K) where the UNM resides \citep[e.g.,][]{McClure-Griffiths2023}.

\begin{figure}
\centering
\includegraphics[width=1.\linewidth]{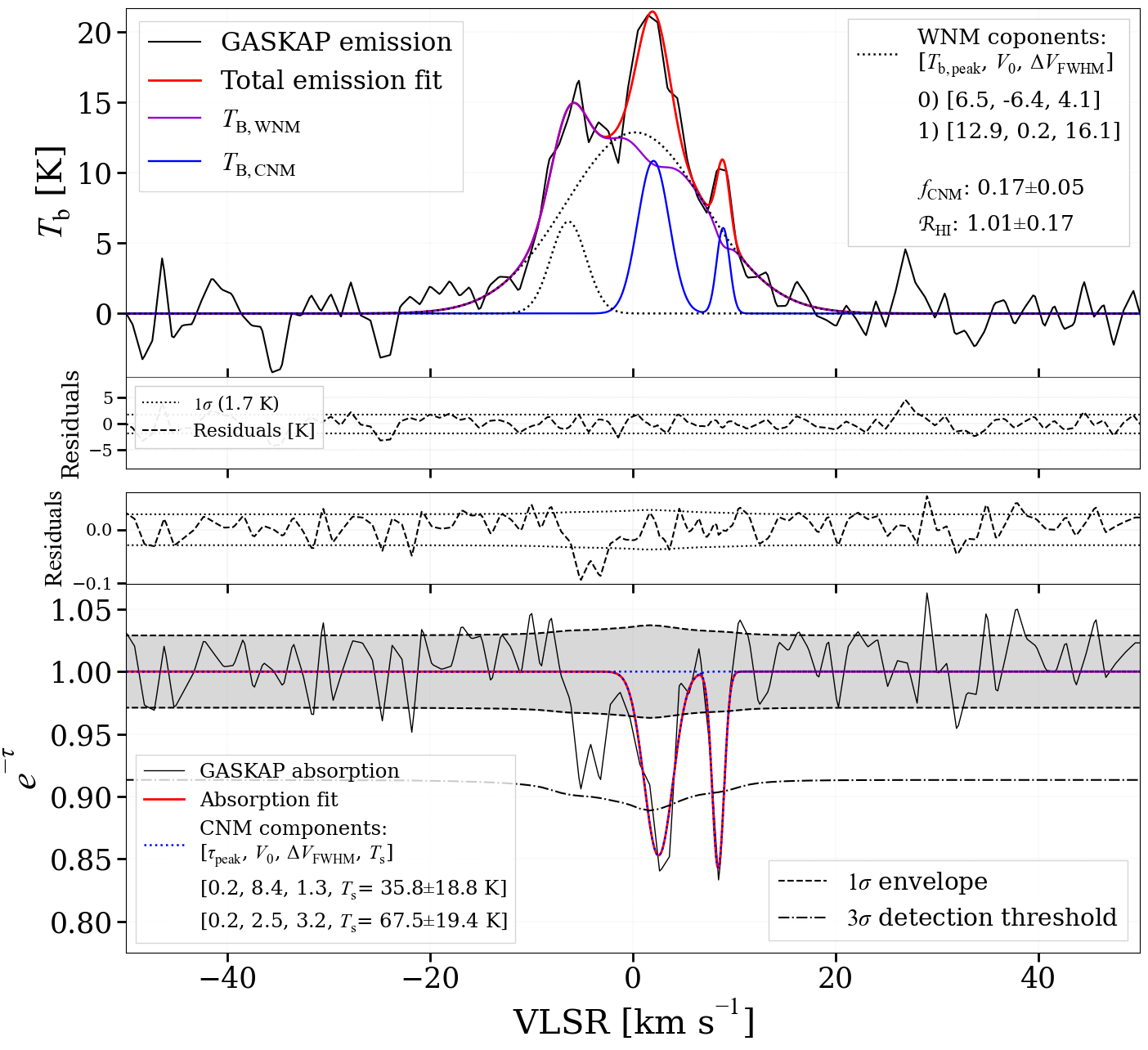}
\caption{Same as Figure \ref{fig:eg_plot1}, but for the source J032548--735326 in SBID 38466. A WNM component ($\Delta V_\mathrm{FWHM}$ = 4.1 \kms, $T_\mathrm{D} \sim 365$ K) is introduced at $-6.4$ \kms\ to account for the associated emission peak, which could be predominantly fulfilled by a potential absorption feature at $-6$ \kms\ that is not identified at the 3$\sigma$ threshold.}
\label{fig:eg_plot2}
\end{figure}

The lowest \TD\ temperatures observed in our WNM components are around 365 K. From inspection of the emission and absorption spectra, there are indications of absorption features at these colder WNM velocities. However, we were unable to detect these absorption features above the 3$\sigma$ detection limit due to the limitations in the sensitivity of our observations. Consequently, the fitting procedure includes a WNM component at these velocities to account for the observed brightness. Figure \ref{fig:eg_plot2} showcases an example where there is likely an absorption feature at $-6$ \kms\ not identified with the 3$\sigma$ threshold, and a WNM component is introduced at $-$6.4 \kms\ to account for the associated emission peak.

We also include the WNM \TD\ histograms, in Figure \ref{fig:ts_td_hist}, from the same sample of high-latitude BIGHICAT sightlines (|$b$| $> 10^{\circ}$, dashed lines) for comparison: orange for the Millennium survey, cyan for the Perseus survey, and blue for 21-SPONGE. The 21-SPONGE \TD\ range from 300 K to 20,000 K ($\Delta V_\mathrm{FWHM} \sim$ 4 \kms\ to 30 \kms), whilst the WNM temperatures for the Millennium and Perseus surveys span an extensive range from 300 K up to 100,000 K ($\Delta V_\mathrm{FWHM} \sim$ 4 \kms\ to 68 \kms). We note that at temperatures above 10,000 K, the WNM gas would mostly become ionized. WNM components with \TD\ $>$ 10,000 K may arise from turbulence within extended WNM regions or could result from multiple blended components along the line of sight (e.g., HT03). The temperatures \Ts\ and \TD\ in three previous works overlap in a range of 300 K to 1000 K (within the UNM regime). The WNM temperature populations in GASKAP-\hi, HT03, and S14 surveys peak at 4000--5000 K, but the 21-SPONGE \TD\ is mostly located in the UNM regime at $\sim$1000 K. The \TD\ CDFs (dashed lines) on the top panel show significant differences in WNM temperatures across surveys, with the exception of the Millennium-Perseus pair (where their K-S test statistic = 0.14, and $p$-value = 0.25).

% See gfit: hi27_.ipynb
% bootstrap(data, errors, num_samples=10000)
% [hnv: Recompute all UNM fractions, using the 250-4000K range!]
Using the \TD\ as an indicator of warm gas kinetic temperatures, we found that 242 out of 570 WNM components (equivalent to $\sim$40\% by column density) are in the thermally unstable regime with \TD\ = 250--4000 K. This implies, on average, 26\% $\pm$ 5\% of the atomic gas in our region of interest does not lie in thermal equilibrium, (where the UNM fraction uncertainty is computed via perturbation bootstrap resampling). Since \TD\ provides an upper limit to the kinetic temperature, this could impact the determination of the UNM fraction, and by extension, the overall fraction of WNM. Our estimated UNM  fraction is nevertheless consistent with previous observational and numerical studies: According to HT03, about 30\% of the total gas is thermally unstable, and \citet{Roy2013} also found at least $\sim$28\% of the gas in the unstable range. The fractions reported in M15 and M18 are slightly lower, with an upper limit of 20\% $\pm$ 10\% derived through calibration of observational methods using numerical simulations. Recent large-scale ISM simulations (TIGRESS-NCR) by \citet{Kim2023TIGRESSNCR} explored different galactic conditions and resolutions for the multiphase ISM and reported a UNM fraction of 27\%–30\%. Similarly, \citet{Bhattacharjee2024} recently analyzed simulations with \hi\ 21 cm observational data in the multiphase neutral ISM and found a UNM fraction (defined as 200 K < \Tk\ < 5000 K, where \Tk\ is the kinetic temperature) of 34\%.

% See gfit/hi20_read_GaussFit_results.ipynb, J032548-73532 in field 38466, use plot_gfit() function

% See hi24_ts_tau.ipynb
\begin{figure}
\includegraphics[width=1.0\linewidth]{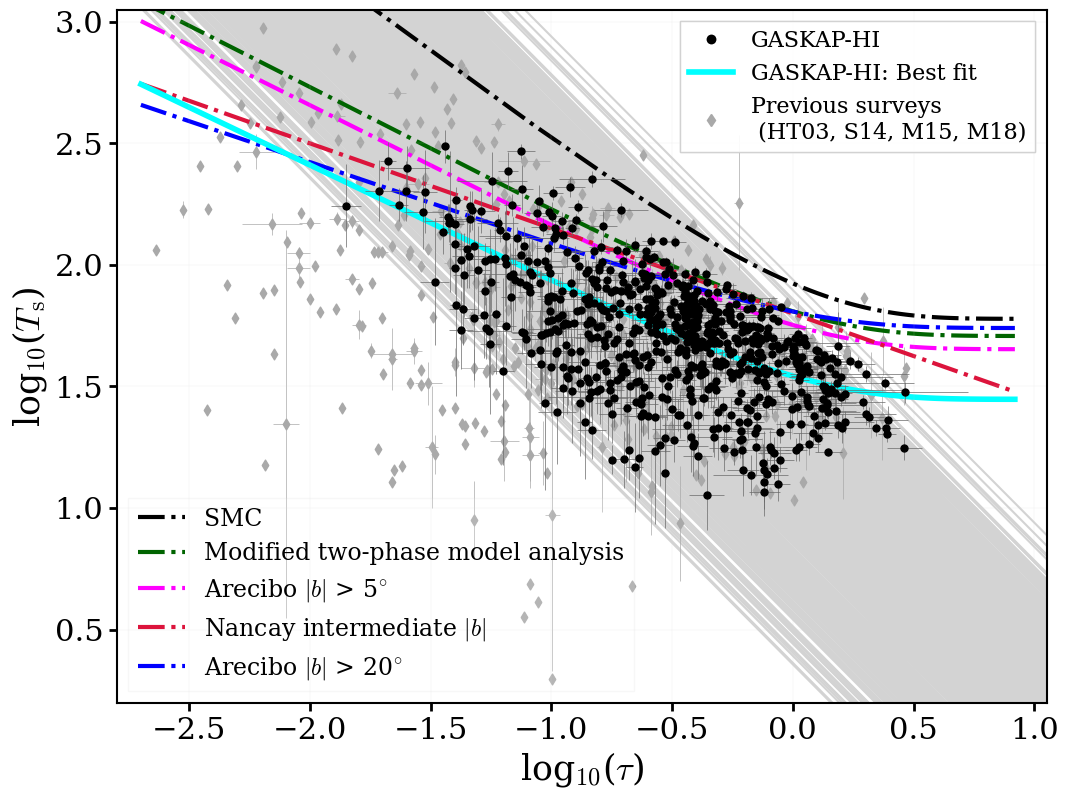}
\caption{The $T_\mathrm{s}$-$\tau$ relationship in log-log space for \ncnm\ absorption Gaussian components detected over \ndet\ lines of sight in current work (black dots). A best fit to GASAKP-\hi\ data is included as a solid cyan line: $T_\mathrm{s} = (28.4\pm1.2) (1-e^{-\tau})^{-(0.48\pm0.02)}$. We draw a gray line through each individual data point using the 21-cm radiative transfer equation (see text for details). The data along 117 high Galactic lines of sight ($|b| > 10^{\circ}$) from previous surveys that used similar Gaussian decomposition techniques (gray diamonds) are also included for comparison: HT03 (51 sightlines), S14 (26 sightlines) and 21-SPONGE (M15 and M18, 40 sightlines). The $T_\mathrm{s}$-$\tau$ curve derived from Arecibo emission and absorption measurements for the SMC by \citealt{Dickey2000} is shown in black, $T_\mathrm{s} = 60 (1-e^{-\tau})^{-0.73}$. The green line corresponds to the numerical analysis of a modified two-phase \hi\ model from \citealt{Liszt1983}, $T_\mathrm{s} = 51 (1-e^{-\tau})^{-0.51}$. The magenta line illustrates a fit to Arecibo emission and absorption measurements toward the intermediate and high Galactic latitudes ($|b| > 5^{\circ}$) by \citealt{Dickey1978}, $T_\mathrm{s} = 45 (1-e^{-\tau})^{-0.5}$. The red line is associated with Nancay absorption measurements at intermediate Galactic latitudes by \citealt{Lazareff1975},$\mathrm{log}_\mathrm{10}T_\mathrm{s} = 1.8 - 0.35 \times \mathrm{log}_\mathrm{10}\tau$. The blue line represents a fit to Arecibo emission and absorption measurements toward high Galactic latitudes ($|b| > 20^{\circ}$) by \citealt{Payne1983}, $T_\mathrm{s} = 55 (1-e^{-\tau})^{-0.34}$.}
\label{fig:ts_vs_tau}
\end{figure}

\subsection{Spin temperature -- optical depth relation}
\label{subsec:ts_tau}

In Figure \ref{fig:ts_vs_tau}, we present in log-log space the relationship between the spin temperature and peak optical depth based on the absorption Gaussian components detected in the current study. The values for the \ncnm\ CNM components, along with their uncertainties, are presented as black data points. From GASKAP emission-absorption Gaussian decomposition, each CNM component has corresponding \Ts, \taupeak, column density (\NHICNM), and line-width ($\Delta V_\mathrm{FWHM}$). Through each individual \Ts-$\tau$ data point, we draw a (gray) line using the 21-cm radiative transfer equation $T_\mathrm{s} = N_{\text{HI,CNM20}} / (0.0195 \times \tau_{\text{peak}} \times \Delta V_\mathrm{FWHM})$ (assuming Gaussian profile, see Equation 8 in \citealt{Heiles2003b} and Equation 1 in \citealt{Dickey1990}) by keeping $N_{\text{HI20,CNM}}$ and $\Delta V_\mathrm{FWHM}$ unchanged but varying the peak optical depths. Here, $N_{\text{HI,CNM20}}$ is the CNM column density expressed in units of $10^{20} \text{ cm}^{-2}$. The \Ts-$\tau$ relations obtained from power law fittings from literature studies are also shown for comparison. These previous works computed the harmonic mean spin temperature $<$$T_\mathrm{s}$$>$ from emission and absorption spectra, rather than decomposing the spectra into Gaussian components. The $T_\mathrm{s}$-$\tau$ curve derived from Arecibo emission-absorption measurements for the SMC by \citealt{Dickey2000} is shown in black. The blue line represents a fit to Arecibo emission and absorption measurements toward high Galactic latitudes ($|b| > 20^{\circ}$) by \citealt{Payne1983}. The green line corresponds to the numerical analysis of a modified two-phase \hi\ model from \citealt{Liszt1983}. The magenta line illustrates a fit to Arecibo emission-absorption measurements toward intermediate and high Galactic latitudes ($|b| > 5^{\circ}$) by \citealt{Dickey1978}. The red line is associated with Nancay absorption measurements at intermediate Galactic latitudes by \citealt{Lazareff1975}.

In the context of \hi\ radiative transfer, the $\tau$-\Ts\ relationship involves interactions among four dependent parameters: column density, spin temperature, optical depth, and line-width (including both thermal and turbulent broadening) as pointed out by \cite{Heiles2003b}. In current work (also in previous absorption studies), the optical depth yet appears to be inversely correlated to the spin temperature of the \hi\ gas, indicating that sightlines with higher optical depth are more likely to be cold and concentrated, leading to lower spin temperatures. Conversely, in regions with lower optical depth, the gas is typically more diffuse and warmer, resulting in higher spin temperatures. This anti-correlation has been a subject of interest in previous studies that utilized Milky Way absorption measurements (intermediate to high Galactic latitudes) and SMC sightlines observed by Arecibo and Nancay radio telescopes \citep[e.g.][]{Lazareff1975,Dickey1978,Liszt1983, Payne1983, Dickey1990, Dickey2000, Heiles2003b}. The outermost gray lines in Figure \ref{fig:ts_vs_tau} determine the lower and upper boundaries for the GASKAP-\hi\ data points. The lower boundary can be expressed as $\mathrm{log}_\mathrm{10}T_\mathrm{s} = 0.3 - \mathrm{log}_\mathrm{10}\tau$, and corresponds approximately to a \Ts-$\tau$ curve of a purely thermally broadened CNM cloud with \NHI\ = 0.01 \nhiUnit\ and $\Delta V_\mathrm{FWHM} = \sqrt{k_\mathrm{B} T_\mathrm{s} / m_\mathrm{H}}$ (the Boltzmann constant constant $k_\mathrm{B}$, and hydrogen atom mass $m_\mathrm{H}$). Whereas, the upper boundary lines up to $\mathrm{log}_\mathrm{10}T_\mathrm{s} = 1.8 - 0.7 \times \mathrm{log}_\mathrm{10}\tau$ curve of a CNM cloud with \NHI\ = 4.7 \nhiUnit, also purely thermally broadened. A least-squares fit to our data yields the best fit: $T_\mathrm{s} = (28.4\pm1.2) (1-e^{-\tau})^{-(0.48\pm0.02)}$. While the \Ts-$\tau$ anti-correlation is expected from \hi\ radiative transfer theory, it may contain fundamental information about the cloud structure and physical conditions within the absorbing regions. The strength and slope of this relationship can reveal important environmental factors, such as variations in density and temperature. For example, \hi\ optical depth may establish the extent to which the spin temperatures can vary. Specifically, the form of the \Ts-$\tau$ relationship means that lower optical depths allow \Ts\ to vary widely, while higher optical depths restrict \Ts\ to a narrower range. For instance, at $\tau <$ 0.25, \Ts\ ranges from 20 K to 300 K; but at $\tau >$ 1.5, \Ts\ is confined to between 20 K and 50 K.

When measurements from previous absorption studies (HT03, S14, M15, and M18) are combined (grey diamonds), the same conclusion may be drawn for the overall \hi\ gas in the local ISM. With the same Solar neighborhood metallicity as the present work, the upper \Ts-$\tau$ boundaries implied by the literature data points remain similar to ours. In contrast, the lower boundary extends to lower $\tau$ values due to the higher optical depth sensitivity. Thus, their \Ts\ range may broaden by up to two orders of magnitude.

Notably, most of our data points lie below the \Ts--$\tau$ relationship observed in previous studies, as seen in Figure \ref{fig:ts_vs_tau}. The discrepancy could mainly be attributed to the methodology used in previous studies, where the harmonic mean spin temperature was derived considering a mixture of both warm and cold gas at the velocity channel of the absorption line (\Ts\ = $T_\mathrm{b}$ / (1 – e$^{-\tau}$) where $T_\mathrm{b}$ is the brightness temperature of the emission line component corresponding to the absorption line). In contrast, our analysis almost directly probes the cold \hi\ gas from the emission-absorption Gaussian decomposition (assuming an isothermal temperature for each Gaussian component).

It is noteworthy that all the Galactic studies lie below the SMC curve from \cite{Dickey2000}. Although the temperature of the CNM clouds in the SMC was found typically at 40 K or less (cooler than the CNM in the local Solar neighborhood), its harmonic mean \hi\ spin temperature at $\sim$440 K appears to be higher than that in the Milky Way. They interpreted this as the typical external pressure in the SMC being much lower than in the Milky Way disk, which cannot confine high-density, cold clouds. Consequently, cold clouds move to pressure equilibrium at lower densities. In addition, photodissociation UV photons may penetrate deeper into the clouds in the SMC than in the Milky Way due to the lower metallicity \citep{Russell1992}, lower dust abundance (less shielding effect) and stronger UV radiation \citep{Rodrigues1997}. As a result, they could find cooler CNM temperatures for \hi\ gas in the SMC, but with a small fraction (less than 15\% of the total \hi) and the remaining WNM making up to 85\%. This mean spin temperature aligns with current SMC studies such as \cite{Dempsey2022} (with an inverse noise-weighted mean spin temperature of \Ts\ $= 245\pm2$ K) and \cite{Jameson2019} (with an arithmetic mean spin temperature of \Ts\ $= 117.2\pm101.7$ K), which analyzed the SMC emission and absorption in similar ways.

\section{Properties along lines of sight}
    \label{results}
    \label{subsec:los_properties}

\subsection{\hi\ column density}
\label{subsec:nhi_dist}

The \hi\ column density of the cold absorbing \hi\ along each sightline is calculated by:

\begin{equation}
N_\mathrm{HI,CNM} = C_0 \int T_\mathrm{s}\ \tau_{v} ~dv,
\label{eq:nhi_cnm}
\end{equation}

\noindent where $C_0 = 1.823\times 10^{18}$ cm$^{-2}$ K$^{-1}$ (km s$^{-1}$)$^{-1}$; for the non-absorbing emission components, we estimate the column density as

\begin{equation}
N_\mathrm{HI,WNM} = C_0 \int T_\mathrm{b}(v) ~dv,
\label{eq:nhi_wnm}
\end{equation}

\noindent where \Tbv\ is the brightness temperature.

\begin{figure}
\includegraphics[width=1.0\linewidth]{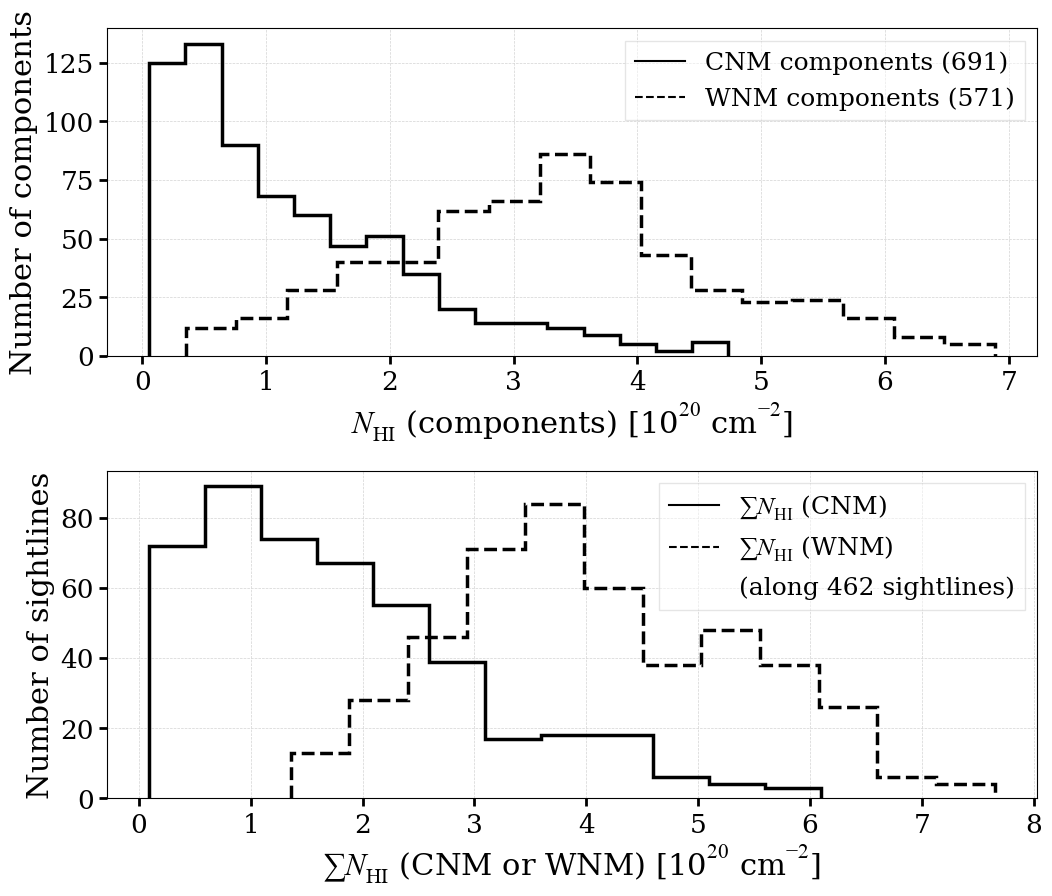}
\caption{Column density histograms of CNM (solid) and WNM (dashed) Gaussian components in the upper panel, $\sum N_\mathrm{HI,CNM}$ and $\sum N_\mathrm{HI,WNM}$ in the lower panel for each line of sight obtained from Gaussian decomposition fit.}
\label{fig:hist_nhi_cpn_los}
\end{figure}

We summarize and highlight in Figure \ref{fig:hist_nhi_cpn_los} the differences between the CNM and WNM column density in the surveyed region. The top panel displays histograms of \NHI\ for individual CNM and WNM Gaussian components. Toward the MC foreground direction, the CNM components cover a column density range from 0.06 \nhiUnit\ to 4.7 \nhiUnit. The distribution is skewed toward lower values, with a mean and median of 1.2 \nhiUnit and 0.9 \nhiUnit, respectively. In contrast, the WNM components have a wider column density range (0.4--6.9) \nhiUnit, and their distribution approaches a Gaussian distribution, with both mean and median around 3.3 \nhiUnit.

In the bottom panel, we show histograms of summed column densities $\sum N_\mathrm{HI,CNM}$ and $\sum N_\mathrm{HI,WNM}$ for CNM and WNM along lines of sight. Similar to the individual components, the summed CNM column density $\sum N_\mathrm{HI,CNM}$ spans a lower range, with a mean of 1.8 \nhiUnit, and a median of 1.6 \nhiUnit, whereas the summed WNM column density $\sum N_\mathrm{HI,WNM}$ spans a higher range, with a mean and median of 4.1 \nhiUnit\ and 3.9 \nhiUnit. The shapes and the column density extents for the summed and individual component values are fairly similar. This is most likely because we detected only about one CNM component and one WNM component per line of sight at these high Galactic attitudes, $b= (-48^{\circ}, -25^{\circ})$. Across all lines of sight, our total CNM column density is about 45\% of the total WNM column density (835.0 \nhiUnit\ compared with 1892.3 \nhiUnit), which yields an average CNM fraction of 28\%$\pm$2\%, where the \FCNM\ uncertainty is obtained by perturbation bootstrap resampling.

% See: hi24_read_results_nhi.ipynb
\begin{figure}
\includegraphics[width=1.0\linewidth]{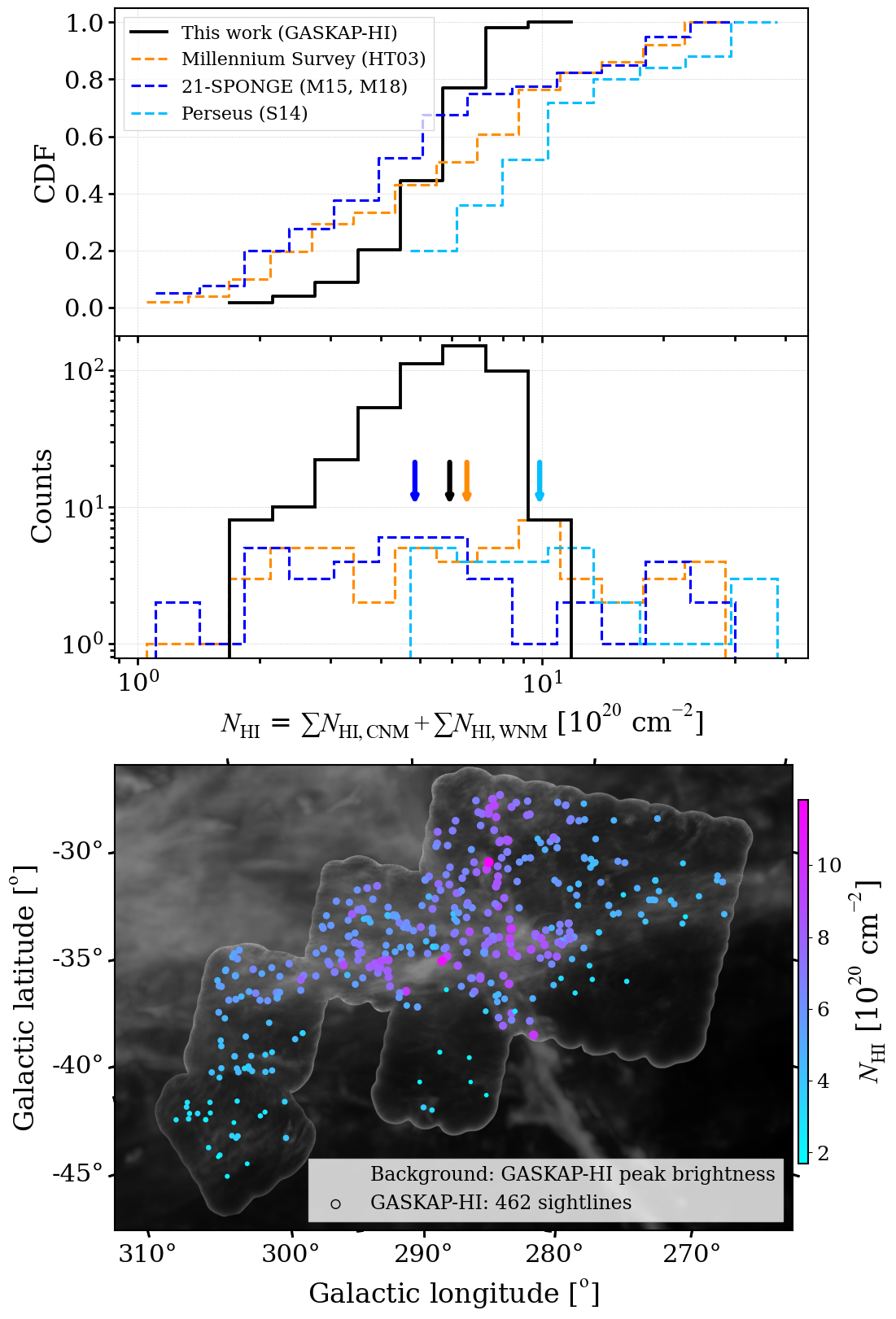}
\caption{Histograms, CDFs, and locations of total \NHI\ = $\sum N_\mathrm{HI,CNM} + \sum N_\mathrm{HI,WNM}$ along \ndet\ GASKAP sightlines (solid lines). The temperatures from a subset of \ntsbiglos\ high-latitude BIGHICAT sightlines ($|b| > 10^\circ$) are shown in the dashed line for comparison: 51 from Millennium survey (HT03, orange), 40 from 21-SPONGE (M15, blue) and 26 from Perseus (S14, cyan). The arrows indicate their median values. In the bottom panel, the GASKAP-\hi\ peak brightness temperature is shown in the gray background (see Figure \ref{fig:all_src_locations} for details); both colors and size represent the GASKAP-\hi\ total \NHI\ magnitudes.}
\label{fig:hist_nhi}
\end{figure}

Figure \ref{fig:hist_nhi} summarizes our findings on the total \hi\ column density along \ndet\ lines of sight, \NHI\ = $\sum N_\mathrm{HI,CNM} + \sum N_\mathrm{HI,WNM}$, and compares them to previous studies at high Galactic latitude $|b| > 10^{\circ}$. The middle panel shows histograms of GASKAP-\hi\ in black, MS in orange, 21-SPONGE in blue, and the Perseus survey in cyan, while the upper panel provides their CDFs. The lower panel shows the total \NHI\ on the GASKAP-\hi\ peak brightness map; both color and sizes of data points denote the column density magnitudes. The median values from the four surveys are indicated by arrows in the middle panel. The Perseus survey, which examined \hi\ gas around the Perseus molecular cloud, reveals the highest \NHI\ range, (5--40) \nhiUnit, with a median of $\sim$10 \nhiUnit. In contrast, the remaining three surveys, which investigate more diffuse Galactic sky, have lower median \NHI\ values.

Our \NHI\ distribution is similar to a bell curve and significantly differs from the previous studies. Namely, the majority of our data points cluster around a narrow band from 1.6 \nhiUnit\ to 11.8 \nhiUnit, with mean and median both at 5.9 \nhiUnit.
% and we do not notice column densities higher than that.
Higher column densities (around 10 \nhiUnit) become apparent near vertical and horizontal filamentary structures, as seen in the lower panel.

So far, we have noticed that the optical depths and spin temperature distribution from HT03 and S14 are likely to be drawn from a similar population; yet, in terms of column density, they are significantly different, as revealed by a K-S test (statistic = 0.4, $p-$value = 0.01). Conversely, another K-S test suggests that the column density distributions of the MS and the 21-SPONGE surveys are not significantly different (K-S statistic=0.2, $p-$value=0.2); thus, their samples may be derived from the same population. This likely reflects genuine differences in their survey targets: the Millennium and 21-SPONGE surveys sampled high-latitude diffuse regions, whereas the S14 sightlines focused on material surrounding a giant molecular cloud.

\subsection{The CNM fraction}
\label{subsec:fcnm}
\begin{figure}
\includegraphics[width=1.0\linewidth]{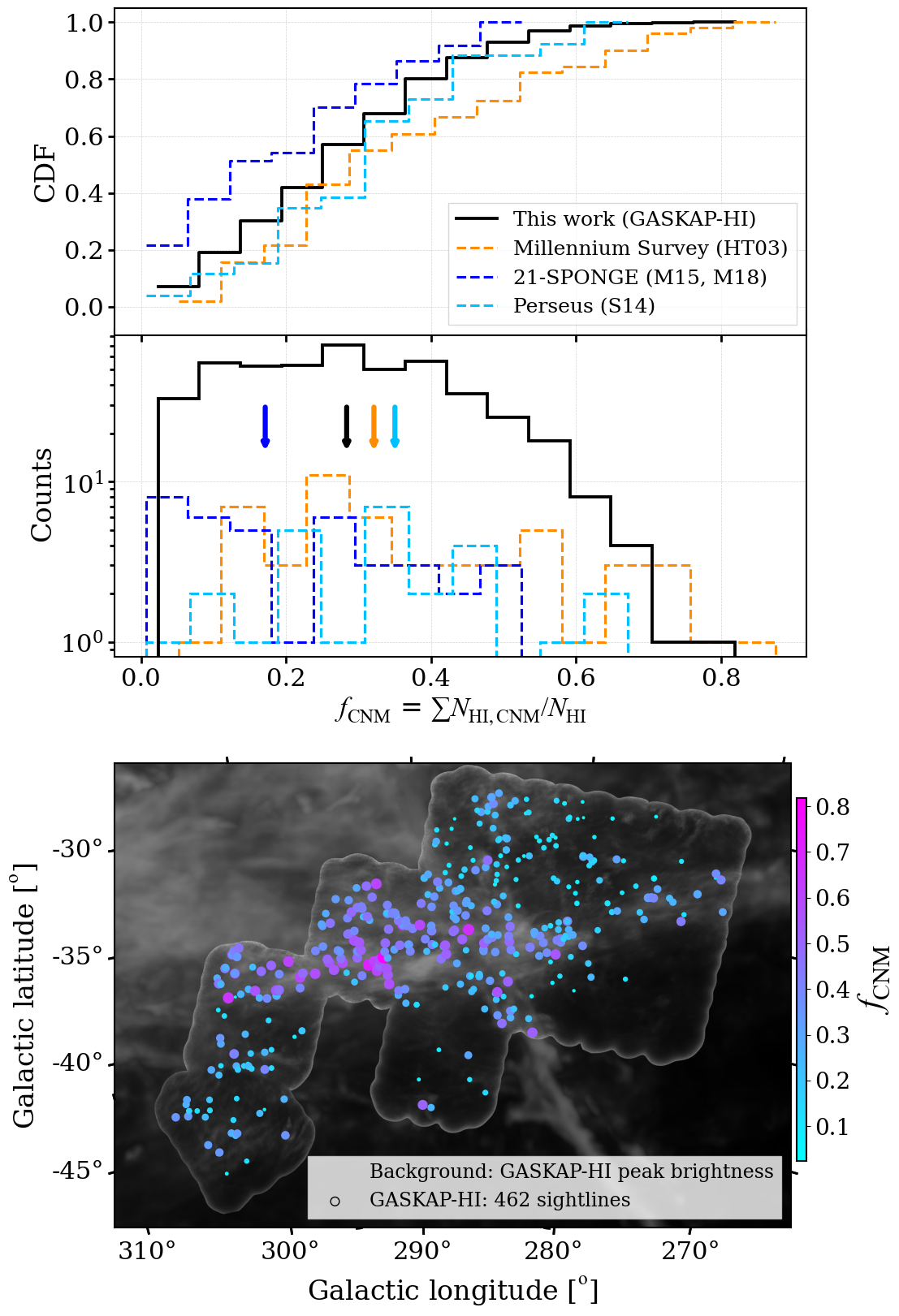}
\caption{Same as Figure \ref{fig:hist_nhi}, but for the CNM fractions $f_\mathrm{CNM}$ = $\sum N_\mathrm{HI,CNM} / N_\mathrm{HI}$.}
\label{fig:fcnm_hist_map}
\end{figure}

Figure \ref{fig:fcnm_hist_map} presents the GASKAP results for the CNM fraction, $f_\mathrm{CNM}$ = $\sum N_\mathrm{HI,CNM} / N_\mathrm{HI}$, and compares them with previous absorption studies. Across all surveys, to ensure a fair comparison of the CNM fraction in this Section, we defined CNM components with \Ts\ $ < 250$ K (in accordance with \citealt{McClure-Griffiths2023}) as cold \hi\ gas. The middle panel displays histograms of \FCNM\ for the four different surveys, black line for the GASKAP-\hi\ (black line), orange line for Millennium Survey, cyan line for Perseus, and blue for 21-SPONGE. The arrows indicate their median values. The upper panel shows their CDFs. The bottom panel depicts the magnitude of the GASKAP cold \hi\ fraction overlaid on the peak brightness map.

Overall, the GASKAP \FCNM\ extends from 0 to 0.82, with both mean and median of (0.29$\pm$0.01). Our \FCNM\ range is comparable to the HT03 Millennium survey, although their mean and median values are slightly higher at 0.32. The S14 Perseus survey, which is directed at the region around the Perseus molecular cloud, has a narrower \FCNM\ range (0 to 0.67), but the highest mean and median of 0.35. The 21-SPONGE survey toward diffuse regions at high galactic latitudes exhibits the narrowest \FCNM\ range and the lowest mean and median values of 0.21 and 0.17, respectively. Based on K-S tests on the \FCNM\ cumulative distributions from the four surveys, the 21-SPONGE survey appears to be significantly different from the others, nonetheless, the GASKAP, Perseus and Millennium \FCNM\ cumulative distributions are likely to come from the same distribution: K-S test statistic=0.22, $p-$value=0.15 for the GASKAP-Perseus pair, and K-S test statistic=0.20, $p-$value=0.45 for the Millennium-Perseus pair. The lower panel shows the \FCNM\ magnitudes overlaid on the GASKAP-\hi\ peak brightness map. The CNM fraction along the \hi\ filaments appears higher than in the surrounding regions. The highest \FCNM\ values ($\sim$0.8) are found mostly along horizontal Hydrus filament.

All Galactic absorption studies, including GNOMES \citep{Nguyen2019} and MACH \citep{Murray2021} surveys, show that the CNM fraction does not exceed $\sim$0.8. This indicates that the WNM is present along all Galactic lines of sight, and it contributes at least 15\% -- 20\% in directions where the CNM dominates. Similar findings were highlighted by S14. The Solar neighborhood CNM fraction is close to the range $\sim$ 40\%--70\% found in the numerical simulations of \citet{Kim2014}.

% See gfit: hi24_read_results_check_beam.ipynb

% See gfit: hi24_read_results_check_beam.ipynb
\begin{figure}
\includegraphics[width=1.0\linewidth]{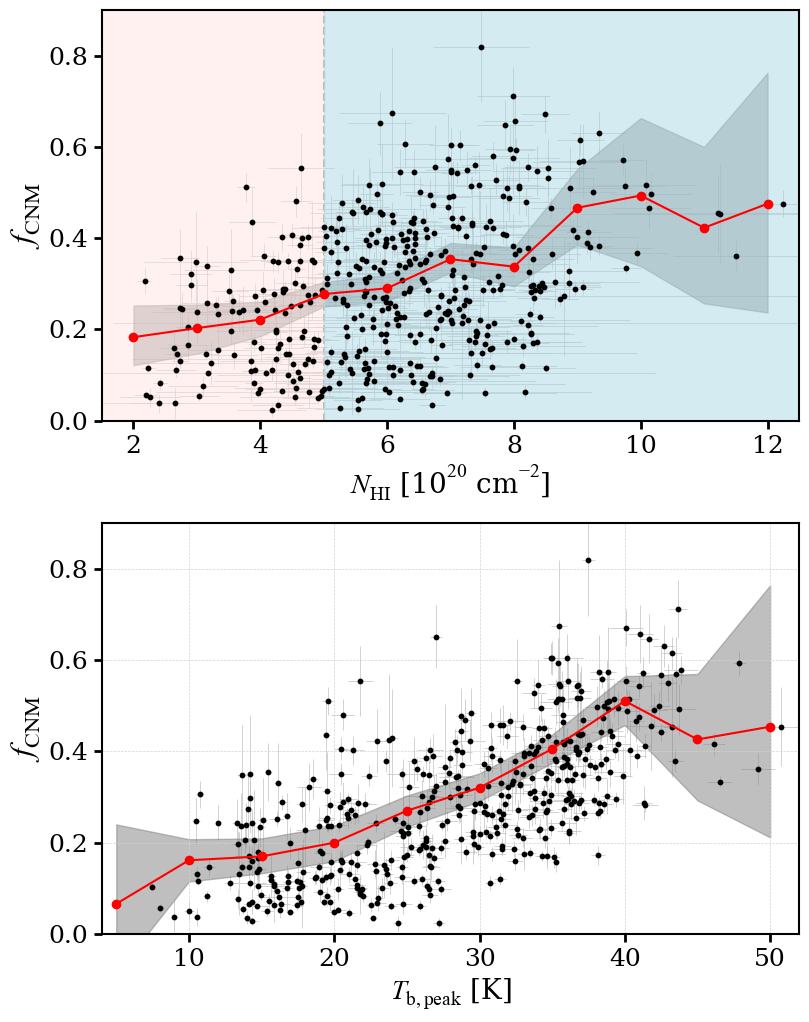}
\caption{The relations between CNM fraction with total \hi\ column density and the peak brightness temperature $T_\mathrm{b, peak}$. Red points represent the medians within bins, while the shaded envelopes show the associated uncertainties, estimated through bootstrap resampling. In the upper panels, the vertical boundary separates the graph with two distinct background colors, positioned at \NHI\ $\sim$ 5 \nhiUnit. 
}
\label{fig:fcnm_nhi_tbpeak}
\end{figure}

In Figure \ref{fig:fcnm_nhi_tbpeak}, we depict how our CNM fraction varies with column density \NHI\ (upper panel) and peak brightness $T_\mathrm{b, peak}$ (lower panel). We computed $T_\mathrm{b, peak}$ as the median of the peak brightness from 20 emission spectra around a radio background source. The red points represent the medians for each \NHI\ or $T_\mathrm{b, peak}$ bins while the shaded envelopes show their associated uncertainties, which were estimated through bootstrap resampling. The \FCNM\ does increase with \NHI\ and peak brightness $T_\mathrm{b, peak}$, yet there is clearly a large scatter. Broadly, the closer to the \hi\ filamentary structures, the higher CNM fractions. 

These trends, where \FCNM\ increases with rising \NHI\ and peak brightness $T_\mathrm{b, peak}$, are consistent with recent studies. Particularly, in an investigation of the physical connection between \hi\ morphology and phase content using Scattering Transform, \citet{Lei2023} found that regions with higher CNM fraction are more populated with small-scale filamentary \hi\ structures. Additionally, \citet{Clark2019} identified that \hi\ filamentary structures in narrow \hi\ channel maps exhibit FWHM line widths, and therefore Doppler temperatures, consistent with CNM temperatures (aligning well with the findings of \citet{Kalberla2016,Kalberla2020}); furthermore, small-scale \hi\ structures in channel maps with increased FIR/\NHI\ ratios are found to originate from a colder, denser phase of the ISM compared to the surrounding material.

\subsection{\hi\ opacity effect}
\label{subsec:R_HI}

In this section, we examine the significance of the \hi\ opacity correction in the Milky Way foreground in front of the LMC and SMC. Through Gaussian decomposition, we obtained the opacity-corrected \hi\ column densities (\NHI). The optically-thin \NHIthin\ is directly estimated from the emission profile, assuming $\tau \ll 1$, making it proportional to the area under the profile. In contrast, calculating the opacity-corrected \NHI\ requires both the spin temperature and optical depth, which we derived from on-/off-source spectra combinations. Our results allow us to compare the two estimates, \NHI\ and \NHIthin\, via their ratio, referred to as the opacity correction factor \RHI = \NHI/\NHIthin.

% See gfit: hi24_read_results_check_beam.ipynb
\begin{figure}
\includegraphics[width=1.0\linewidth]{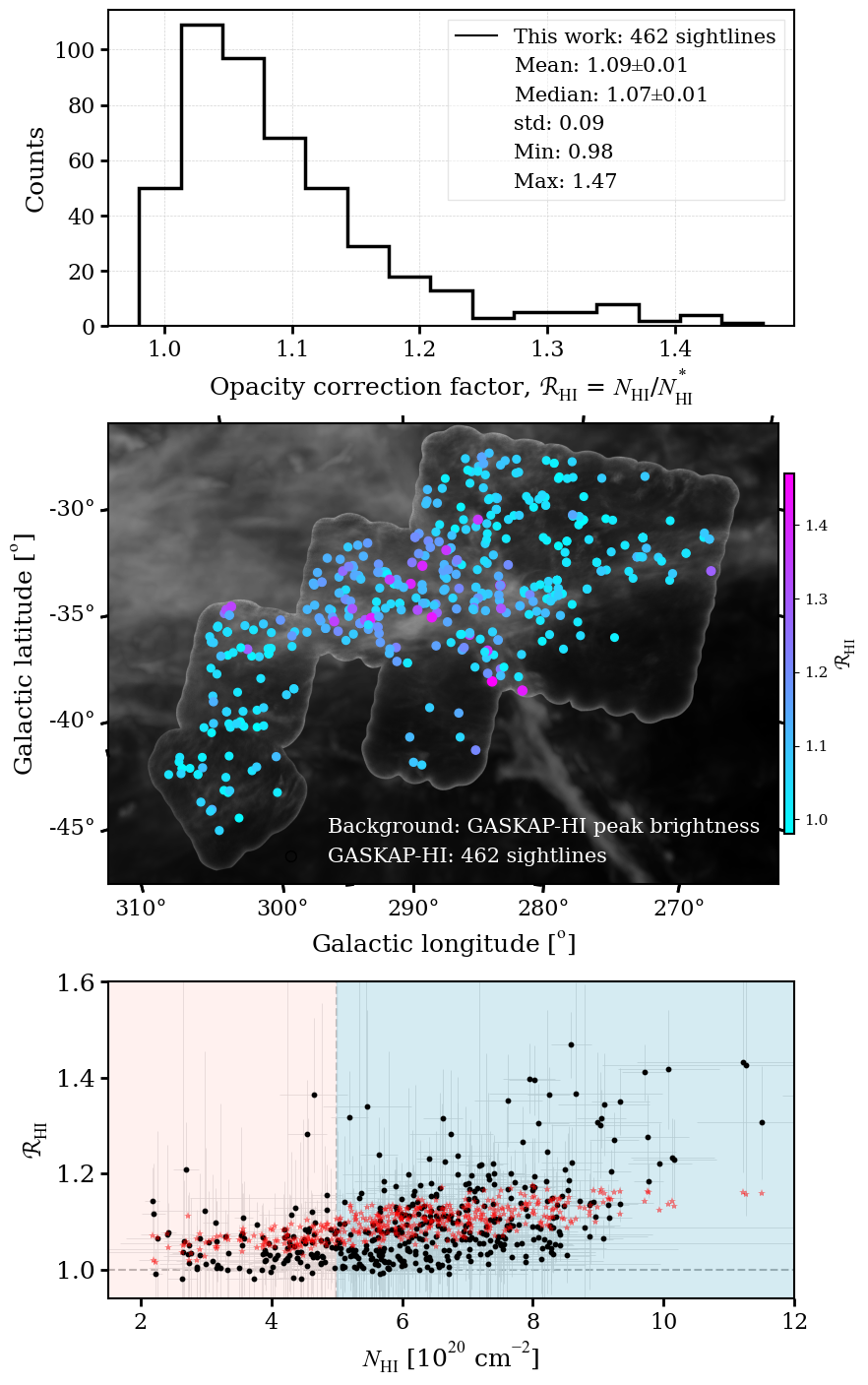}
\caption{Top panel: Histogram of \RHI = \NHI/\NHIthin\ in the MC foreground. Middle panel: The GASKAP-\hi\ peak brightness temperature is shown in the gray background (see Figure \ref{fig:all_src_locations} for details); both colors and sizes represent the ratio \RHI\ magnitudes. Bottom panel: \RHI = \NHI/\NHIthin\ as a function of \NHI\ in the MC foreground. The vertical dashed line represents \NHI\ = 5 \nhiUnit, and the horizontal dashed line indicates the points where \NHI\ = \NHIthin. Red stars show corrected column density $N_\mathrm{HI,iso}$ assuming a single isothermal representative spin temperature \Ts\ $\sim$ 102 K for the entire survey region.}
\label{fig:rhi_hist_map}
\end{figure}

Figure \ref{fig:rhi_hist_map} presents our \RHI\ histogram (upper panel), \RHI\ magnitudes, and their relationship with column density for \ndet\ GASKAP-\hi\ sightlines. Over the full sample, the \RHI\ factor varies from $\sim$1.0 to 1.4, with a mean and median of (1.09$\pm$0.01) and (1.07$\pm$0.01), respectively. Most of the \RHI\ values are slightly above unity. The relationship between \RHI\ and \NHI\ in the lower panel indicates that opacity correction becomes important at higher \hi\ column densities. Specifically, for low column densities below 5$\times 10^{20}$ cm$^{-2}$ (light-pink slice in lower panel), \RHI\ is consistent with unity within the uncertainties, and \NHIthin\ is thus comparable to the corrected \NHI. As the column density increases, the ratio rises, reaching \RHI\ $\sim1.3$ when the \NHI\ doubles to $\sim$10$\times 10^{20}$ cm$^{-2}$ (light-blue shaded area). The \RHI\ and \NHI\ relation reveals an overall increase. The opacity effect increases with greater column densities of the two vertical and horizontal filaments, particularly near their intersection, as displayed in the middle panel. Overall, with \RHI\ $\sim$ 1.1 in the directions of the Magellanic Cloud foreground, the correction for opacity effects thus appears to be small.

On a sightline-by-sightline basis, we perform a simple correction to the optically-thin assumption using a single temperature, $T_\mathrm{s,iso}$, for the entire observing area. The isothermal opacity-corrected column density can be estimated from the emission profile and an isothermal spin temperature:

\begin{equation}
N_\mathrm{HI,iso} = 1.8224 \times 10^{18} \int T_\mathrm{s} ~ ln \left( \frac{T_\mathrm{s}}{T_\mathrm{s} - T_\mathrm{b}(v)} \right) dv
\label{eq_nhi_iso}
\end{equation}

\noindent where $T_\mathrm{b}(v)$ is brightness temperature (see \citealt{Lockman1995,Wakker2011}). In our diffuse high Galactic latitudes, this approach is feasible due to low optical depth, which prevents \Ts\ from approaching $T_\mathrm{b}$ and the denominator from being zero. A representative spin temperature $T_\mathrm{s,iso}$ $\approx$ 102 K provides a good agreement with the GASKAP data, shown as red stars in Figure \ref{fig:hist_nhi}. Previous literature values of \RHI\ include $\sim$ (1.1--1.3) for 79 random sightlines in the Millennium survey; \RHI\ $\sim$ (1.0--1.1) for 40 high Galactic latitude lines of sight in the 21-SPONGE survey; and in the range of (1.0-1.8) toward 77 lines of sight in the vicinity of Taurus and Gemini molecular clouds \citep{Nguyen2019}.

We have observed about ten low column density lines of sight with \RHI\ factors around 0.98 instead of the expected value of 1.0, but consistent with unity within uncertainties. These minor discrepancies can be attributed to slight underfitting of the emission profiles, which have significant noise.

% \section{Properties along filaments}
%     \label{filaments}
%     \input{6_filaments}

\section{\hi\ vs dust}
    \label{hi_dust}
    \label{sec:hi_dust}

% See gfit: Radiation03_thermal_dust.ipynb
\begin{figure}
\includegraphics[width=1.0\linewidth]{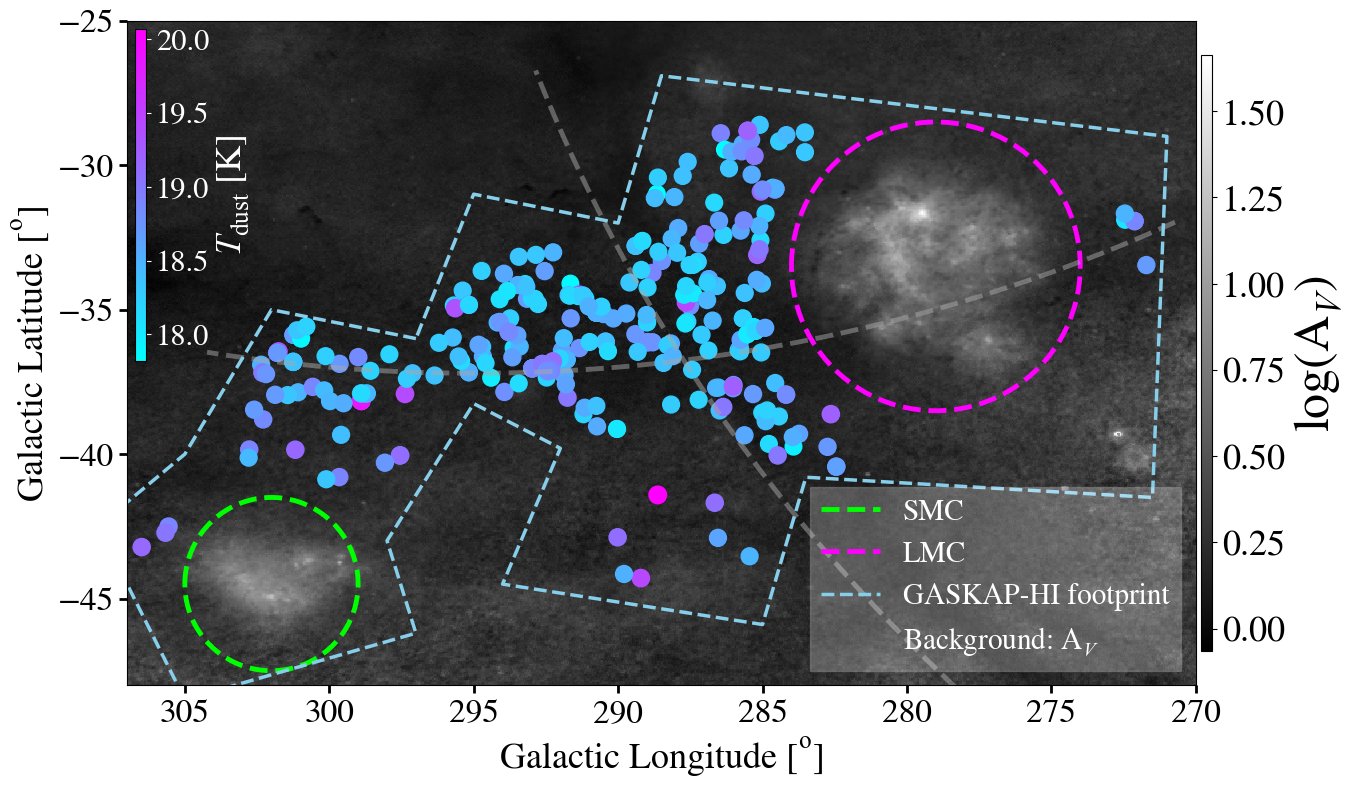}
\caption{Locations of 290 lines of sight far away from the directions to the SMC (green circle) and LMC (magenta circle). The background represents the dust visual extinction $A_\mathrm{V}$ inferred from the \textit{Planck} dust reddening \ebv: \av\ = $R_\mathrm{V}$ \ebv, with an average ``ratio of total to selective extinction'' $R_\mathrm{V} = 3.1$. The colors of data points illustrate dust temperatures $T_\mathrm{dust}$. The dashed light-blue boundary shows the approximate GASKAP-\hi\ observing footprint.}
\label{fig:dust_los}
\end{figure}

% See Table 1: https://www.aanda.org/articles/aa/pdf/2010/15/aa15441-10.pdf
In this section, we investigate the relations between the \hi\ gas and dust properties in the direction of MC foreground. We use the reddening \ebv\ map produced by \cite{Schlegel1998}, which is based on dust FIR observations from COBE/DIRBE and IRAS \citep{Silverberg1993,Neugebauer1984}. This map covers the entire sky with an angular resolution of 6$^{\prime}$.1 (a linear size of $\sim$0.4 pc, assuming a distance $d =$ 250 pc to the \hi\ clouds in our region of interest). \citet{Schlafly2011} revisited the calibration of the SFD98 map and found that the original SFD98 map overestimated \ebv\ by approximately 14\%. To address this, they performed a detailed recalibration of the SFD98 map using photometric data from the Sloan Digital Sky Survey and provided correction factors for various photometric bands.  For the work presented here, we apply the correction from Table 6 of \citet{Schlafly2011} to convert SFD98 reddening to dust visual extinction (centered at $\lambda$ = 551 nm): \av\ = 2.742 $E(B-V)_\mathrm{SFD98}$, with an average ``ratio of total to selective extinction'' \rv\ = 3.1 for the Galactic ISM. To estimate the uncertainties in reddening values, we rely on reddening prediction accuracy of 16\%, as stated in SFD98. In addition, we employ $Planck$ thermal dust maps \citep{PLC2014} with angular resolution is 5$^{\prime}$ (a linear size of $\sim$0.3 pc), which is comparable to the effective angular distances between GASKAP emission and absorption (around 1$^{\prime}$ -- 2$^{\prime}$, corresponding to a linear size of 0.7 -- 0.14 pc). \citet{PLC2014} provided all-sky maps of dust optical depth \t353, dust temperature $T_\mathrm{dust}$, dust spectral index $\beta_\mathrm{dust}$, and their corresponding uncertainties. These parameters were derived from modified blackbody fits to \textit{Planck} dust intensities at 353, 545, and 857 GHz, as well as IRAS 100 $\mu$m using a single dust temperature component along a line of sight. Since the SFD98 and $Planck$ dust maps sample the full sightlines, including dust emission in the SMC and LMC, we exclude sightlines close to the SMC and LMC to avoid contamination from their dust content. This exclusion leaves 290 lines of sight for the current analysis. Figure \ref{fig:dust_los} outlines the locations of the 290 selected lines of sight on the \textit{Planck} dust visual extinction map \av, with colors indicating dust temperatures $T_\mathrm{dust}$. The \textit{Planck} dust visual extinction $A_\mathrm{V}$ is inferred from their dust reddening \ebv\, which is estimated based on a model of dust emission, and also provided by \citet{PLC2014}. We then use their \t353-\ebv\ correlation, \ebv/\t353\ $= (1.49 \pm 0.03) \times 10^{4}$, to estimate the uncertainty in \ebv, and subsequently in \av.

\subsection{ \hi\ gas vs dust extinction}
\label{subsec:nh-av}

% See gfit: Radiation03_thermal_dust.ipynb
% See gfit: dr3dust01_NHI_Av_Planck.ipynb
% See gfit: dr3dust02_NHI_Av.ipynb
\begin{figure}
\includegraphics[width=1.0\linewidth]{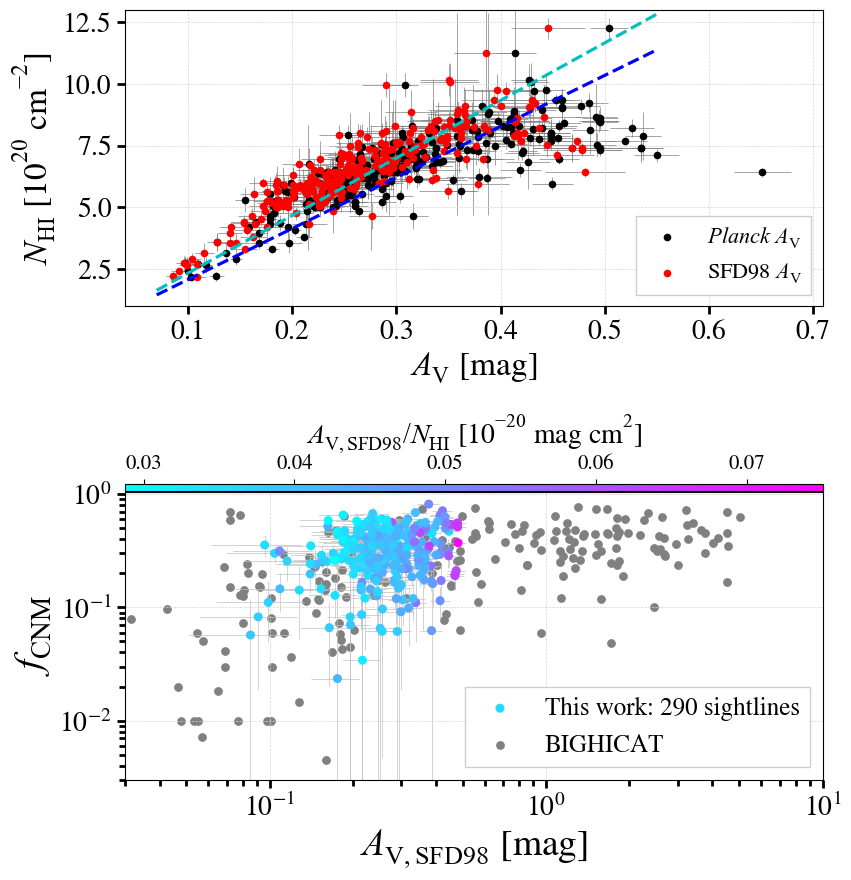}
\caption{Upper panel: Relation between dust visual extinction \av\ and total \hi\ column density \NHI\ for a sample of 290 lines of sight far away from the directions to the SMC and LMC. Black data points represent \textit{Planck} \av, red points for the \av\ from SFD98. The blue and cyan lines indicate the linear fits for \textit{Planck} \av\ and the SFD98 \av, respectively. Lower panel: CNM fraction vs \NHI\ for the same 290 lines of sight (color-coded by the magnitude of $A_\mathrm{V,SFD98}/N_\mathrm{HI}$. The values along 373 BIGHICAT sightlines are shown (in gray) for comparison. 
}
\label{fig:nhi_fcnm_av}
\end{figure}
% extinction_curve.txt
% https://zenodo.org/records/7811871

In Figure \ref{fig:nhi_fcnm_av}, we assess the impact of dust extinction on \hi\ column density and CNM fraction. The top panel shows the total \hi\ column density plotted against SFD98 \av\ as red points. This reveals an overall increasing trend of \NHI\ with increasing \av\ from 0.09 mag to 0.48 mag, with greater scatter at higher \av. A linear fit \NHI/$10^{20}$ cm$^{-2}$ $= (23.32 \pm 0.22)$ \av\ (cyan dashed line) to the data points indicates that the trend deviates from linearity as \NHI\ and \av\ rise, showing an excess of dust extinction. With \rv = 3.1 for the Galactic ISM, the excess of dust extinction translates into an excess of dust reddening \ebv\ = \av/\rv\ at higher \hi\ column densities, which agrees with the findings from \cite{Lenz2017}.

In addition to SFD98 data, we also extract the \av\ values from \citet{PLC2014} for our \ndet\ lines of sight. The \textit{Planck} visual extinction is presented in black dots along with a linear fit \NHI/$10^{20}$ cm$^{-2}$ $= (20.68 \pm 0.23)$ \av\ (blue dashed line). The \textit{Planck} \av\ was inferred from thermal dust emission, which is mainly dependent on the temperatures, composition, sizes, and emissivity of dust gains. The \textit{Planck} \av\ ranges from 0.12--0.59 mag (mean and median at 0.30 mag, respectively), meanwhile SFD98 \av\ spans a narrower band from $\sim$0.09 to 0.48 mag (both mean and median at 0.26 mag). In relation to total \NHI\, \textit{Planck} visual extinction exhibits larger scatter across the entire range, and indicates higher dust excess at high gas column densities.

Interstellar dust can influence the thermal balance of the ISM, so regions with high dust content may have conditions that favor the formation of CNM, potentially a higher CNM fraction \citep{Murray2020cnn}. In the bottom panel in Figure \ref{fig:nhi_fcnm_av}, we present the changes in \FCNM\ with visual extinction \av. Our data points represent a sample towards 290 Magellanic foreground lines of sight and are color-coded by the ratio $A_\mathrm{V,SFD98}/N_\mathrm{HI}$, while the gray dots in the background correspond to data from previous absorption surveys (entire BIGHICAT sample of 373 lines of sight). Altogether, existing absorption surveys encompass a wide range of local Galactic ISM conditions with \av\ from 10$^{-3}$ to 10 mag. Although it overall appears that higher visual extinction leads to a higher cold \hi\ gas fraction, our data do not exhibit a clear trend between \FCNM\ and \av, mainly because we are probing a diffuse region within a narrow range of \av\ (0.1 -- 0.6 mag). Our sample, as expected, highlights a distinct gradient of the ratio $A_\mathrm{V}/N_\mathrm{HI}$ in the \FCNM-\av\ space, with a higher CNM fraction associated with greater dust excess. This result is in excellent agreement with the findings of \citet{Murray2020cnn}, who observed that the ratio between dust intensity at 857 Ghz and optically-thin column density \( I_{857}/N^{*}_{\text{HI}} \) increases with rising \FCNM. It is also consistent with \citet{Clark2019}, who demonstrated that \( I_{857}/N^{*}_{\text{HI}} \) increases with small-scale structure intensity. These findings collectively suggest that the observed small-scale structures are preferentially comprised of CNM.

Towards sightlines with \av\ below 0.2 mag the CNM contributes slightly to the total column density, \FCNM\ $\lesssim$ 10\%. However, as soon as the \av\ reaches 0.2 mag, the CNM fraction spans a broad range from $\sim$10\% up to 80\%, with most values around 30\%. Then beyond \av\ $\approx$ 0.6 mag, indicating denser directions, about 50\% of \hi\ gas is found in the cold phase. In the local ISM, the extinction threshold for \h2\ hydrogen molecules to form is \av\ $\gtrsim$ 0.14 mag (CO requires \av\ $\gtrsim$ 0.8 mag, \citealt{Wolfire2010}). OH molecules have been detected at \av\ $\approx$ 0.1 mag \citep{Li2018}, and \citet{Rybarczyk2022} found an \av\ threshold of $\sim$0.25 mag for the existence of HCO$^{+}$ in the diffuse ISM. This indicates that the transition from WNM to CNM and then atomic to molecular gas can occur at \av\ $\sim$ 0.1 mag in the diffuse Galactic ISM. This \av\ threshold for the transition to molecules is also in phase with the theoretical work of \citet{Federman1979}.

\subsection{Radiation field vs \hi\ temperature}

% See gfit: Radiation03_thermal_dust.ipynb
\begin{figure}
\includegraphics[width=1.0\linewidth]{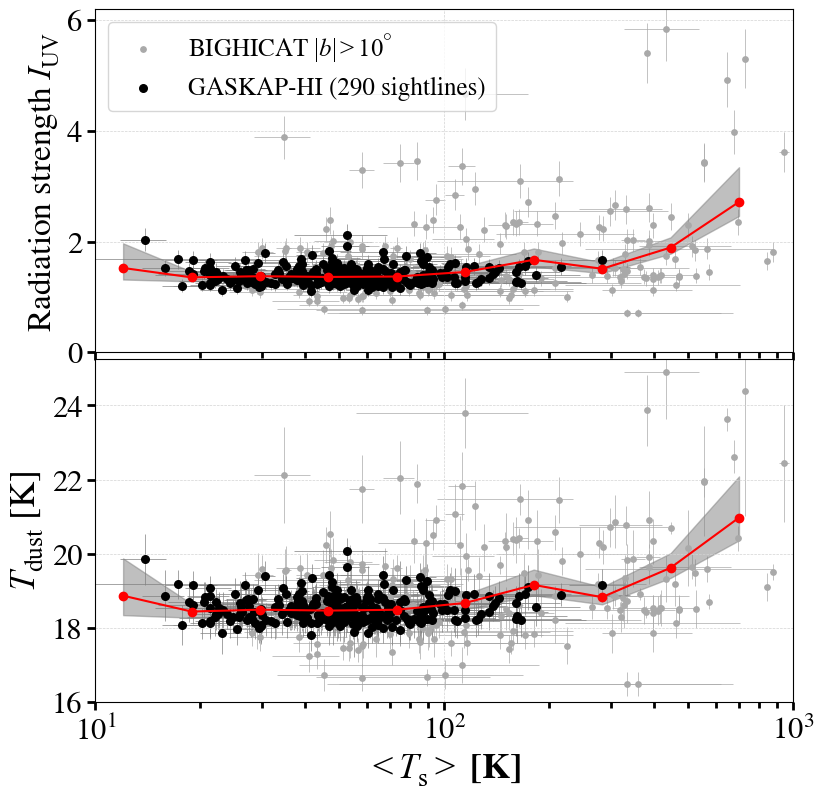}
\caption{CNM column density-weighted mean temperatures $<$$T_\mathrm{s}$$>$ as a function of dust proxies for radiation field: estimates of UV radiation strength $I_\mathrm{UV}$ (top panel) and dust temperature $T_\mathrm{dust}$ (bottom panel). Black data points are from GASKAP-\hi\ (excluding sightlines towards the SMC and LMC), gray points are from 214 BIGHICAT lines of sight with $|b| > 10^{\circ}$. Red points represent the medians within $<$$T_\mathrm{s}$$>$ bins for both GASKAP-\hi\ (290 black points) and BIGHICAT (214 gray points) data, with shaded envelopes indicating the associated uncertainties estimated through bootstrap resampling.} 
\label{fig:ts_mean_vs_dust}
\end{figure}

In this section, we investigate how the radiation field affects \hi\ gas temperatures. Neutral gas temperatures are determined by heating and cooling processes. The radiation field plays a key role in heating and thus affects the distribution of \hi\ thermal phases. Understanding that our survey, which focuses on a small region in front of the SMC and LMC, has a narrow temperature range, we incorporate previous absorption studies (214 lines of sight from BIGHICAT with $|b| > 10^{\circ}$) for various local ISM lines of sight to explore a broader range of temperatures and radiation field strengths.

We employ \textit{Planck} thermal dust maps to estimate the radiation field through two methods: Firstly, dust temperature serves as a first-order approximation of the ambient radiation field. Higher dust temperatures generally indicate regions with more intense heating. Secondly, we estimate the UV radiation strength for dust grains at high Galactic latitudes $|b| > 10^{\circ}$ \citep{Boulanger1996,Paradis2011} relative to the Draine field \citep{Draine1978,Bialy2020,Park2023} using $I_\mathrm{UV} = (T_\mathrm{d} / 17.5)^{\beta + 4}$, where typical solar neighborhood conditions give $I_\mathrm{UV} \sim 1$.

Figure \ref{fig:ts_mean_vs_dust} illustrates the variation of the column density-weighted mean CNM temperature $<$$T_\mathrm{s}$$>$ with these two proxies for the ambient radiation field strength: $I_\mathrm{UV}$ (top panel), and $T_\mathrm{d}$ (bottom panel). Red points represent the medians of $<$$T_\mathrm{s}$$>$ values within bins for both GASKAP-\hi\ (black points) and BIGHICAT (gray points) data, while the shaded envelopes show the associated uncertainties estimated through bootstrap resampling. In both cases, we do not observe a significant variation in our $<$$T_\mathrm{s}$$>$ with radiation field strength, leaving mostly flat distributions for the GASKAP-\hi\ temperatures ranging from $\sim$10 to 300 K. However, previous high-sensitivity absorption surveys (HT03, S14, M18) have recorded higher \hi\ gas temperatures (up to 1000 K) at higher dust temperatures (up to 22 K -- 24 K) and stronger radiation fields (e.g., $I_\mathrm{UV} \sim 4$ Draine field units). Although there is considerable scatter and fewer measurements at higher temperatures, the combined trends shown in Figure \ref{fig:ts_mean_vs_dust} deliver tentative evidence that stronger interstellar radiation fields may result in higher \hi\ temperatures. More measurements of higher temperatures (at high optical depth sensitivities) are required to confirm this trend.

% \section{\hi\ emission at high and low resolutions}
%     \label{res_comparision}
%     \input{8_res_comparison}

\section{Conclusions and Future work}
    \label{conclusions}
    \label{sec:conclusions}
In this study, we have conducted an extensive analysis of the local  neutral ISM using the largest Galactic \hi\ absorption survey with the Australian SKA Pathfinder Telescope (GASKAP-\hi) at unprecedented high spatial resolution (30$^{\prime\prime}$). We aimed to investigate the physical properties of cold and warm \hi\ gas in the Solar neighborhood at high Galactic latitudes ($-45^{\circ}, -25^{\circ}$) toward the Magellanic Cloud foreground. The surveyed area encompasses the intersection of two prominent \hi\ filaments: the vertical Reticulum filament and the horizontal Hydrus filament (see Figure \ref{fig:all_src_locations}).

Through the GASKAP-\hi\ survey, we measured the \hi\ absorption along \nlos\ continuum background sources, achieving $\sim$12 absorption measurements per square degree. This density of measurements, made possible by the GASKAP large field of view, provides a dense grid of \hi\ absorption lines of sight, approaching the density of emission measurements (though still far from equal). We detected strong \hi\ absorption at the 3$\sigma$ threshold towards \ndet\ background radio continuum sources, yielding an absorption detection rate of 17\% and two detections per square degree.

The unprecedented 30$^{\prime\prime}$ angular resolution, corresponding to a linear scale of 0.04 pc (assuming distance to gas clouds $d = 250$ pc), enables GASKAP-\hi\ to measure emission extremely close to the absorption (effective angular distance of $\sim$45$^{\prime\prime}$), ensuring that the emission and absorption sample similar \hi\ parcels along a given sky direction. By performing joint Gaussian decompositions using the method developed by \cite{Heiles2003a} for all pairs of GASKAP absorption-emission spectra, we directly determined \hi\ optical depths, temperatures, and column densities for \ncnm\ CNM and 570 WNM Gaussian components. We then examine the correlation between CNM spin temperatures and optical depths, and the dust-gas relationship in Magellanic Clouds' foreground. Our key conclusions are as follows:

$-$ The \hi\ gas in our surveyed area at high Galactic latitudes predominantly exists in a diffuse atomic state, with a median \taupeak\ $\sim$ 0.35, \Ts\ $\sim$ 50 K for CNM, and a \Tkmax\ $< 20,000$ K for WNM (Sections \ref{subsec: optical_depth} and \ref{subsec:temperatures}).

$-$ The thermal properties of cold \hi\ gas in the Magellanic Cloud foreground are in excellent agreement with those observed in previous absorption surveys under various local ISM conditions: random sky directions, in the vicinity of the Perseus molecular cloud, near the anti-Galactocentric giant molecular clouds or within the high-density Riegel-Crutcher cloud. These findings provide evidence that the CNM properties are fairly prevalent throughout the local ISM (see Section \ref{subsec:temperatures}), consistent with the conclusions drawn by S14 and \cite{Nguyen2019}.

$-$ We noticed an anti-correlation between CNM spin temperature and optical depth in the Magellanic Cloud foreground (see Section \ref{subsec:ts_tau}). %Higher optical depth tends to exhibit lower spin temperatures. 
The range of possible spin temperatures in thermal equilibrium is likely determined by \hi\ optical depth (which is primarily a function of density). Lower optical depths allow \Ts\ to vary widely, whereas higher $\tau$ results in a narrower range of \Ts. Including findings from previous high-sensitivity absorption observations, a similar conclusion can be drawn for the CNM across the local Galactic ISM.

$-$ Along the \ndet\ GASKAP-\hi\ Galactic sightlines with absorption detections, on average, about 30\% of neutral gas is in the CNM phase. The \FCNM\ increases from 0 to 0.82 with increasing \NHI\ and brightness \TBpeak. Generally, regions closer to \hi\ filaments exhibit higher CNM fractions (Section \ref{subsec:fcnm}).

$-$ Our findings reveal a general linear relation between \hi\ column density and dust visual extinction \av, but an excess of dust \av\ at higher column density, in agreement with previous results \citep{PLC2014,Lenz2017}. Although higher visual extinction corresponds to a higher fraction of cold \hi\ gas \citep{McClure-Griffiths2023}, our data do not show a clear trend between the \FCNM\ and \av\ within the narrow dust extinction range of our diffuse high-latitude regions. The Galactic radiation field in front of SMC and LMC is relatively uniform, resulting in flat distributions when compared to GASKAP mean spin temperatures. Nevertheless, data from previous high-sensitivity absorption studies (HT03, S14, M15, M18) suggest a hint of higher \hi\ spin temperatures in regions with a stronger radiation field.

Our future work will focus on analyzing the distribution and structures of \hi\ gas in the Magellanic Cloud foreground where the two prominent filaments reside (Lynn et al. 2024b, prep.). By comparing the current GASKAP-\hi\ emission and absorption datasets with the latest 3D dust maps \citep{Edenhofer2024}, we aim to obtain a 3D view of gas and dust of these filamentary structures. Additionally, still using GASKAP-\hi\ absorption, we have carried out joint emission-absorption decomposition with GASS emission data measured at lower angular resolution (16$^{\prime}$). We will compare these results with the current fitting outcomes to examine the influence of spatial resolutions on the derived cold \hi\ properties. The ongoing GASKAP-\hi\ main survey, which involves 20 times longer integration times and significantly improved brightness and optical depth sensitivities, will enable the detection of absorption components at lower optical depths and higher temperatures. This will enhance our ability to study the transition of the warm neutral medium to unstable neutral medium and cold neutral medium.

\section*{Acknowledgements}
This scientific work uses data obtained from Inyarrimanha Ilgari Bundara / the Murchison Radio-astronomy Observatory. We acknowledge the Wajarri Yamaji People as the Traditional Owners and native title holders of the Observatory site. CSIRO’s ASKAP radio telescope is part of the Australia Telescope National Facility \hyperlink{https://ror.org/05qajvd42}{(https://ror.org/05qajvd42)}. Operation of ASKAP is funded by the Australian Government with support from the National Collaborative Research Infrastructure Strategy. ASKAP uses the resources of the Pawsey Supercomputing Research Centre. Establishment of ASKAP, Inyarrimanha Ilgari Bundara, the CSIRO Murchison Radio-astronomy Observatory and the Pawsey Supercomputing Research Centre are initiatives of the Australian Government, with support from the Government of Western Australia and the Science and Industry Endowment Fund.

This research was partially funded by the Australian Government through an Australian Research Council Australian Laureate Fellowship (project number FL210100039) to NMc-G. SS acknowledges the support provided by the University of Wisconsin-Madison
Office of the Vice Chancellor for Research and Graduate Education with funding from the Wisconsin Alumni Research Foundation, and the NSF Award AST-2108370. JDS acknowledges funding by the European Research Council via the ERC Synergy Grant ``ECOGAL -- Understanding our Galactic ecosystem: From the disk of the Milky Way to the formation sites of stars and planets'' (project ID 855130). SEC acknowledges support from NSF award AST-2106607 and an Alfred P. Sloan Research Fellowship. Finally, we thank the anonymous referee for the comments and suggestions that allowed us to improve the quality of our manuscript.

\textit{Software}: Astropy \citep{Pytorch2019}, Matplotlib \citep{MatplotlibHunter2007}, NumPy \citep{vanderWalt2011}, SciPy \citep{Virtanen2020}, Pandas \citep{mckinney2010data}.
% \citep{Astropy2018A}, PyTorch 

%%%%%%%%%%%%%%%%%%%%%%%%%%%%%%%%%%%%%%%%%%%%%%%%%%
\section*{Data Availability}
This paper includes archived data obtained through the CSIRO ASKAP Science Data Archive, CASDA \href{https://research.csiro.au/casda}{(https://research.csiro.au/casda)}.

The GASKAP emission and absorption data used in this study, along with the fitted results and their associated uncertainties are available at \href{https://github.com/GASKAP/HI-Absorption/tree/master/MW\_absorption}{GASKAP Github repository}.



\bibliographystyle{mnras}
\bibliography{references} % if your bibtex file is called example.bib

%\begin{thebibliography}{99}
%\end{thebibliography}

%%%%%%%%%%%%%%%%%%%%%%%%%%%%%%%%%%%%%%%%%%%%%%%%%%

%%%%%%%%%%%%%%%%% APPENDICES %%%%%%%%%%%%%%%%%%%%%

\appendix

% \section{Gaussian decomposition with single dish emission data}

% Here...

%%%%%%%%%%%%%%%%%%%%%%%%%%%%%%%%%%%%%%%%%%%%%%%%%%

% Don't change these lines
\bsp	% typesetting comment
\label{lastpage}
\end{document}